\documentclass[10pt,fleqn, table]{article}

\usepackage[english]{babel}
\usepackage[utf8]{inputenc}
\usepackage[T1]{fontenc}
\usepackage{lmodern}

\usepackage{natbib}
\bibpunct{(}{)}{,}{a}{,}{,}

\usepackage{amsmath, amssymb, amsfonts, amsthm}

\usepackage{cancel}
\usepackage{mathrsfs}
\usepackage{dsfont}

\usepackage{fleqn}
\numberwithin{equation}{section}

\usepackage{ragged2e}
\usepackage{soul}
\usepackage{textcomp}

\usepackage{graphicx}
\usepackage[format=hang, font=small, labelfont=bf]{caption}
\usepackage{tikz}
\usetikzlibrary{decorations.pathreplacing}
\usetikzlibrary{positioning, calc}

\usepackage{array}
\usepackage{colortbl}
\usepackage{multirow, multicol}
\usepackage{color}
\definecolor{gris50}{gray}{0.50}
\definecolor{gris95}{gray}{0.95}
\definecolor{gris85}{gray}{0.85}
\definecolor{mygreen}{rgb}{0.16,0.5,0.32}
\definecolor{myblue}{RGB}{51,102,153}

\usepackage{tabularx}
\usepackage[algo2e,ruled]{algorithm2e}
\usepackage{setspace}

\usepackage{enumitem}

\usepackage{listings}
\usepackage[a4paper,citecolor=blue,linkcolor=black,colorlinks=true, urlcolor=blue]{hyperref}

\usepackage{chngpage}
\usepackage{pdfpages}
\usepackage{authblk}

\usepackage{geometry}

\usepackage{fancyhdr}
\newcommand{\knowing}{\;\ifnum\currentgrouptype=16 \middle\fi|\;}
\newcommand{\suchthat}{\;\ifnum\currentgrouptype=16 \middle\fi|\;}

\newcommand{\s}{\mathscr{S}}
\newcommand{\G}{\mathscr{G}}
\newcommand{\ivj}{i\stackrel{\G}{\sim} j}

\newcommand{\N}{\mathscr{N}}
\newcommand{\C}{\mathscr{C}}
\newcommand{\Ss}{\mathbf{S}}

\newcommand{\Xv}{\mathbf{X}}
\newcommand{\x}{\mathbf{x}}
\newcommand{\X}{\mathscr{X}}

\newcommand{\Y}{\mathscr{Y}}
\newcommand{\y}{\mathbf{y}}
\newcommand{\Yv}{\mathbf{Y}}
\newcommand{\yobs}{\mathbf{y}^{\text{obs}}}

\newcommand{\step}{\epsilon}

\newcommand{\pot}{\mathrm{V}}

\newcommand{\btheta}{\boldsymbol\theta}

\newcommand{\bTheta}{\boldsymbol\Theta}
\newcommand{\bphi}{\boldsymbol\phi}
\newcommand{\bpsi}{\boldsymbol\psi}
\newcommand{\bPsi}{\boldsymbol\Psi}

\newcommand{\pr}{\mathbf{P}}
\newcommand{\esp}{\mathbf{E}}
\newcommand{\var}{\textbf{Var}}
\newcommand{\cov}{\textbf{Cov}}
\newcommand{\ind}{\mathbf{1}}
\DeclareMathOperator*{\argmin}{arg\,min\,}
\DeclareMathOperator*{\argmax}{arg\,max\,}

\DeclareMathOperator*{\hess}{\mathbf{\nabla}^2}
\newcommand{\grad}{\mathbf{\nabla}}

\newcommand{\D}{\mathcal{D}}
\newcommand{\M}{\mathscr{M}}

\newcommand{\vertiii}[1]{{\left\vert\kern-0.25ex\left\vert\kern-0.25ex\left\vert #1 
    \right\vert\kern-0.25ex\right\vert\kern-0.25ex\right\vert}}
\newcommand{\KL}{\text{KL}}

\newcommand{\dd}{\mathrm{d}}

\newcommand{\eg}{\textit{e.g.},~}
\newcommand{\ie}{\textit{i.e.},~}

\newcommand{\mle}{\text{MLE}}

\newcommand{\mfl}{\text{MF-like}}

\newcommand{\BIC}{\text{BIC}}

\begin{document}
\title{A review on statistical inference methods for discrete Markov random fields}

\author[1]{Julien Stoehr}

\affil[1]{School of Mathematical Sciences \& Insight Centre for Data Analytics, University College Dublin, Ireland}
\date{}
\maketitle

\begin{abstract}
Developing satisfactory methodology for the analysis of Markov random field is a very challenging task. Indeed, due to the Markovian dependence structure, the normalizing constant of the fields cannot be computed using standard analytical or numerical methods. This forms a central issue for any statistical approach as the likelihood is an integral part of the procedure. Furthermore, such unobserved fields cannot be integrated out and the likelihood evaluation becomes a doubly intractable problem. This report gives an overview of some of the methods used in the literature to analyse such observed or unobserved random fields.

\vspace{0.5cm} \noindent \textbf{Keywords:} statistics; Markov random fields; parameter estimation; model selection.
\end{abstract}

\section{Introduction}

The problem of developing satisfactory methodology for the analysis of spatially correlated data has been of a constant interest for more than half a century now. Constructing a joint probability distribution to describe the global properties of such data is somewhat complicated but the difficulty can be bypassed by specifying the local characteristics via conditional probability instead. This proposition has become feasible with the introduction of Markov random fields (or Gibbs distribution) as a family of flexible parametric models for spatial data \citep[\textit{the Hammersley-Clifford theorem}, ][]{besag1974}. Markov random fields are spatial processes related to lattice structure, the conditional probability at each nodes of the lattice being dependent only upon its neighbours, that is useful in a wide range of applications. In particular, hidden Markov random fields offer an appropriate representation for practical settings where the true state is unknown. The general framework can be described as an observed data $\y$ which is a noisy or incomplete version of an unobserved discrete latent process $\x$.

Gibbs distributions originally appears in statistical physics to describe equilibrium state of a physical systems which consists of a very large number of interacting particles such as ferromagnet ideal gases \citep{lanford1969}. But they have since been useful in many other modelling areas, surged by the development in the statistical community since the 1970's. Indeed, they have appeared as convenient statistical model to analyse different types of spatially correlated data. Notable examples are the autologistic model \citep{besag1974} and its extension the Potts model. Shaped by the development of \cite{geman1984} and \cite{besag1986}, -- see for example \citet{alfo2008} and \citet{moores2014} who performed image segmentation with the help of this modelling --  and also in other applications including disease mapping \citep[\textit{e.g., }][]{green2002} and genetic analysis \citep[\textit{e.g., }][]{francois2006, friel2009} to name a few. The exponential random graph model or $p^{\ast}$ model \citep{wasserman1996} is another prominent example \citep{frank1986} and arguably the most popular statistical model for social network analysis \citep[\textit{e.g., }][]{robins2007}.

Interests in these models is not so much about Markov laws that may govern data but rather the flexible and stabilizing properties they offer in modelling. 
Whilst the Gibbs-Markov equivalence provides an explicit form of the joint distribution and thus a global description of the model, this is marred by a considerable computational curse. Conditional probabilities can be easily computed, but the joint and the marginal distribution are meanwhile unavailable
since the normalising constant is of combinatorial complexity and generally can not be evaluated with standard analytical or numerical methods. This forms a central issue in statistical analysis as the computation of the likelihood is an integral part of the procedure for both parameter inference 
and model selection. 
Remark the exception of small latices on which
we can apply the recursive algorithm of \cite{reeves2004, friel2007} and obtain an exact computation of the normalizing constant. However, the complexity in time of the aforementioned algorithm grows exponentially and is thus helpless on large lattices. Many deterministic or stochastic approximations have been proposed for circumventing this difficulty and developing methods that are computationally efficient and accurate is still an area of active research. Solutions to deal with the intractable likelihood are of two kinds. On one hand, one can rely on pseudo-model as surrogates for the likelihood. Such solutions typically stems from composite likelihood or variational approaches. On the other hand Monte Carlo methods have played a major role to estimate the intractable likelihood in both frequentist and Bayesian paradigm.


The present survey paper cares about the problem of carrying out statistical inference (mostly in a Bayesian framework) for Markov random fields. When dealing with hidden random fields, the focus is solely on hidden data represented by discrete models such as the Ising or the Potts models. Both are widely used examples and representative of the general level of difficulty. Aims may be to infer on parameters of the model or on the latent state $\x$. The paper is organised as follows: it begins by introducing the existence of Markov random fields with some specific examples (Section \ref{sec:markov-gibbs}). The difficulties inherent to the analysis of such a stochastic model are especially pointed out in Section \ref{sec:issues}. The special case of small regular lattices is described in Section \ref{sec:recursive}. As befits a survey paper, Section \ref{sec:pseudo-model} focuses on solutions based on pseudo-models while Section \ref{sec:sampling-method} is dedicated to a state of the art concerning sampling method.

\section{Markov random field and Gibbs distribution}
\label{sec:markov-gibbs}

\subsection{Gibbs-Markov equivalence}

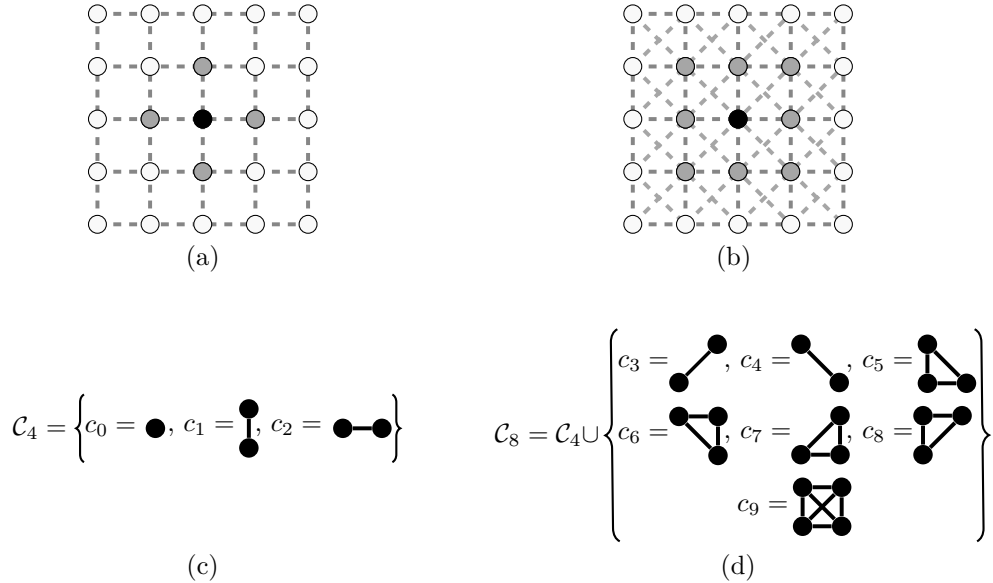
\begin{figure}[t]
\centering
\begin{minipage}[t]{7cm}
\centering
\begin{tikzpicture}[scale=0.7]
\foreach \x in {0,...,4}{
  \draw[dashed, line width=1.5pt, gray!95] (\x,0) to[out=90,in=-90] (\x,4);
  \draw[dashed, line width=1.5pt, gray!95] (0,\x) to[out=0,in=180] (4,\x);
}

\foreach \x in {0,...,4} 
	\foreach \y in {0,...,4}
   		\draw[fill = gray!5] (\x,\y) circle (1.7mm); 

\draw[fill = black] (2,2) circle (1.7mm); 
\draw[fill = gray!70] (2,3) circle (1.7mm); 
\draw[fill = gray!70] (1,2) circle (1.7mm); 
\draw[fill = gray!70] (3,2) circle (1.7mm); 
\draw[fill = gray!70] (2,1) circle (1.7mm); 

\end{tikzpicture}\\
(a)
\end{minipage}
\vspace{0.5cm}
\begin{minipage}[t]{7cm}
\centering
\begin{tikzpicture}[scale=0.7]
\foreach \x in {0,...,4}{
  \draw[dashed, line width=1.5pt, gray!95] (\x,0) to[out=90,in=-90] (\x,4);
  \draw[dashed, line width=1.5pt, gray!95] (0,\x) to[out=0,in=180] (4,\x);
}

\foreach \x in {1,...,3}{
  \draw[dashed, line width=1.5pt, gray!70] (0,\x) to[out=-45,in=135] (\x,0);
  \draw[dashed, line width=1.5pt, gray!70] (\x,4) to[out=-45,in=135] (4,\x);
  \draw[dashed, line width=1.5pt, gray!70] (\x,0) to[out=45,in=225] (4,4-\x);
  \draw[dashed, line width=1.5pt, gray!70] (0,\x) to[out=45,in=225] (4-\x,4);
}

\draw[dashed, line width=1.5pt, gray!70] (0,0) to[out=45,in=225] (4,4);
\draw[dashed, line width=1.5pt, gray!70] (0,4) to[out=-45,in=135] (4,0);

\foreach \x in {0,...,4} 
	\foreach \y in {0,...,4}
   		\draw[fill = gray!5] (\x,\y) circle (1.7mm); 

\draw[fill = black] (2,2) circle (1.7mm); 
\draw[fill = gray!70] (2,3) circle (1.7mm); 
\draw[fill = gray!70] (1,2) circle (1.7mm); 
\draw[fill = gray!70] (3,2) circle (1.7mm); 
\draw[fill = gray!70] (2,1) circle (1.7mm); 
\draw[fill = gray!70] (1,3) circle (1.7mm); 
\draw[fill = gray!70] (3,3) circle (1.7mm); 
\draw[fill = gray!70] (1,1) circle (1.7mm); 
\draw[fill = gray!70] (3,1) circle (1.7mm); 

\end{tikzpicture}\\
(b)
\medskip
\end{minipage}
\\
\begin{minipage}[t]{7cm}
\centering
\begin{tikzpicture}[scale=1]

\node[circle, fill = black, scale = 0.78] (first) at (0.,0.){};
\node[left = -0.12cm of first] (lfirst) {$c_0 =\,\,$};

\node[right = -0.1cm of first] (second) {, $c_1 =\,\,$};
\node[circle, fill = black, scale = 0.78, above right = -0.1cm of second, xshift = -0.1cm] (tsec) {}; 
\node[circle, fill = black, scale = 0.78, below = 0.25cm of tsec] (bsec) {}; 
\draw[line width=1.5pt] (tsec) to[out=-90,in=90] (bsec);

\node[right = -0.1cm of second] (third) {, $c_2 =\,\,$};
\node[circle, fill = black, scale = 0.78, right = -0.1cm of third] (ater) {}; 
\node[circle, fill = black, scale = 0.78, right = 0.25cm of ater] (bter) {}; 
\draw[line width=1.5pt] (ater) to[out=0,in=180] (bter);

\draw [decorate, decoration = {brace,mirror, amplitude=3pt}, xshift = -27pt, yshift = 0pt, thick, left = of lfirst]
(0.,0.45) -- (0.,-0.45) node [midway,xshift=-.1cm] 
{$\mathcal{C}_4 =$};

\draw [decorate, decoration = {brace, amplitude=3pt}, xshift = -0pt, yshift = 0pt, thick, right = of bter]
(3.15,0.45) -- (3.15,-0.45);

\node at (0, -1.4){};

\end{tikzpicture}\\
(c)
\end{minipage}
\begin{minipage}[t]{7cm}
\centering
\begin{tikzpicture}[scale=1]

\node (first) at (0,0) {$c_3 =\,$};
\node[circle, fill = black, scale = 0.78, below right = -0.1cm of first, xshift = -0.1cm] (afirst) {}; 
\node[right = 0.25cm of afirst] (bfirst) {}; 
\node[circle, fill = black, scale = 0.78, above = 0.25cm of bfirst] (cfirst) {}; 
\draw[line width=1.3pt] (afirst) to[out=45,in=225] (cfirst);

\node[right = 0.42cm of first] (sec) {, $c_4 =\,$};
\node[circle, fill = black, scale = 0.78, above right = -0.1cm of sec, xshift = -0.1cm] (bsec) {}; 
\node[below = 0.25cm of bsec] (asec) {}; 
\node[right = 0.25cm of bsec] (dsec) {}; 
\node[circle, fill = black, scale = 0.78, right = 0.25cm of asec] (csec) {}; 
\draw[line width=1.5pt] (bsec) to[out=-45,in=135] (csec);

\node[right = 0.42cm of sec] (third) {, $c_5 =\,\,$};
\node[circle, fill = black, scale = 0.78, above right = -0.1cm of third, xshift = -0.1cm] (bthird) {}; 
\node[circle, fill = black, scale = 0.78, below = 0.25cm of bthird] (athird) {}; 
\node[circle, fill = black, scale = 0.78, right = 0.25cm of athird] (cthird) {}; 
\draw[line width=1.5pt] (bthird) to[out=-90,in=90] (athird);
\draw[line width=1.5pt] (athird) to[out=0,in=180] (cthird);
\draw[line width=1.5pt] (cthird) to[out=135,in=-45] (bthird);

\node[below = 0.5cm of first] (four) {$c_6 =\,$};
\node[circle, fill = black, scale = 0.78, above right = -0.1cm of four, xshift = -0.1cm] (afour) {}; 
\node[circle, fill = black, scale = 0.78,  right = 0.25cm of afour] (bfour) {}; 
\node[circle, fill = black, scale = 0.78, below = 0.25cm of bfour] (cfour) {}; 
\draw[line width=1.5pt] (afour) to[out=0,in=180] (bfour);
\draw[line width=1.5pt] (bfour) to[out=-90,in=90] (cfour);
\draw[line width=1.5pt] (cfour) to[out=135,in=-45] (afour);

\node[right = 0.42cm of four] (five) {, $c_7 =\,$};
\node[circle, fill = black, scale = 0.78, below right = -0.1cm of five, xshift = -0.1cm] (afive) {}; 
\node[circle, fill = black, scale = 0.78, right = 0.25cm of afive] (bfive) {}; 
\node[circle, fill = black, scale = 0.78, above = 0.25cm of bfive] (cfive) {}; 
\draw[line width=1.5pt] (afive) to[out=0,in=180] (bfive);
\draw[line width=1.5pt] (bfive) to[out=90,in=-90] (cfive);
\draw[line width=1.5pt] (cfive) to[out=225,in=45] (afive);

\node[right = 0.42cm of five] (six) {, $c_8 =\,$};
\node[circle, fill = black, scale = 0.78, below right = -0.1cm of six, xshift = -0.1cm] (asix) {}; 
\node[circle, fill = black, scale = 0.78, above = 0.25cm of asix] (bsix) {}; 
\node[circle, fill = black, scale = 0.78, right = 0.25cm of bsix] (csix) {}; 
\draw[line width=1.5pt] (asix) to[out=90,in=-90] (bsix);
\draw[line width=1.5pt] (bsix) to[out=0,in=180] (csix);
\draw[line width=1.5pt] (csix) to[out=225,in=45] (asix);

\node[below = 0.5cm of five, xshift = 0.1cm] (sev) {$c_9 =\,\,$};
\node[circle, fill = black, scale = 0.78, below right = -0.1cm of sev, xshift = -0.1cm] (asev) {}; 
\node[circle, fill = black, scale = 0.78, right = 0.25cm of asev] (bsev) {}; 
\node[circle, fill = black, scale = 0.78, above = 0.25cm of bsev] (csev) {}; 
\node[circle, fill = black, scale = 0.78, left = 0.25cm of csev] (dsev) {};
\draw[line width=1.5pt] (asev) to[out=0,in=180] (bsev);
\draw[line width=1.5pt] (bsev) to[out=90,in=-90] (csev);
\draw[line width=1.5pt] (csev) to[out=180,in=0] (dsev);
\draw[line width=1.5pt] (dsev) to[out=-90,in=90] (asev);
\draw[line width=1.5pt] (csev) to[out=225,in=45] (asev);
\draw[line width=1.5pt] (bsev) to[out=135,in=-45] (dsev);

\draw [decorate, decoration = {brace, amplitude=5pt}, xshift = -25pt, yshift = 0pt, thick]
($(bsev) + (-2.95, -0.2)$) -- ($(dsec) + (-2.95, 0.2)$) node [midway,xshift=-0.95cm] 
{$\mathcal{C}_8 = \mathcal{C}_4 \cup$};

\draw [decorate, decoration = {brace, mirror, amplitude=5pt}, xshift = -0pt, yshift = 0pt, thick]
($(bsev) + (1.8, -0.2)$) -- ($(dsec) + (1.8, 0.2)$);

\end{tikzpicture}\\
(d)
\end{minipage}
\caption{\small First and second order neighbourhood graphs $\G$ with corresponding cliques. (a) The four closest neighbours graph $\G_4$. neighbours of the vertex in black are represented by vertices in gray. (b) The eight closest neighbours graph $\G_8$. neighbours of the vertex in black are represented by vertices in gray. (c) Cliques of graph $\G_4$. (d) Cliques of graph $\G_8$.}
\label{fig:neigh}
\end{figure}

A discrete random field $\Xv$ is a collection of random variables $X_i$ indexed by a finite set  $\s = \{1,\dots,n\}$, whose elements are called sites, and taking values in a finite state space $\mathscr{X}:=\{0,\ldots,K-1\}$. For a given subset $A\subset\s$, $\Xv_A$ and $\x_A$ respectively define the random process on $A$, \ie, $\{X_i, i\in A\}$, and a realization of $\Xv_A$. Denote by $\s\setminus A= -A$ the complement of $A$ in $\s$. When modelling local interactions, the sites are lying on an undirected graph $\G$ which induces a topology on $\s$: by definition, sites $i$ and $j$ are adjacent or neighbour if and only if $i$ and $j$ are linked by an edge in $\G$. A random field $\Xv$ is a Markov random field with respect to $\G$, if for all configuration $\x$ and for all sites $i$ it satisfies the following Markov property
\begin{equation}
\pr\left(X_i = x_i\suchthat \Xv_{-i} = \x_{-i}\right) = \pr\left(X_i = x_i\suchthat \Xv_{\N(i)} = \x_{\N(i)}\right),
\label{eqn:markov}
\end{equation}
where $\N(i)$ denotes the set of all the adjacent sites to $i$ in $\G$.
It is worth noting that any random field is a Markov random field with respect to the trivial topology, that is the cliques of $\G$ are either the empty set or the entire set of sites $\s$. Recall a clique $c$ in an undirected graph $\G$ is any single vertex or a subset of vertices such that every two vertices in $c$ are connected by an edge in $\G$. As an example, when modelling a digital image, the lattice is interpreted as a regular 2D-grid of pixels and the random variables states as shades of grey or colors. Two widely used adjacency structures are the graph $\G_4$ (first order lattice), respectively $\G_8$ (second order lattice), for which the neighbourhood of a site is composed of the four, respectively eight, closest sites on a two-dimensional regular lattice, except on the boundaries of the lattice, see Figure \ref{fig:neigh}.

The difficulty with the Markov formulation is that one defines a set of conditional distributions which does not guarantee the existence of a joint distribution. The Hammersley-Clifford theorem states that if the distribution of a Markov random field with respect to a graph $\G$ is positive for all configuration $\x$ then it admits a Gibbs representation for the same topology (see \eg \cite{grimmett1973, besag1974} and for a historical perspective \cite{clifford1990}),  
namely a density function on $\X$ parametrised by $\btheta\in\bTheta\subset\mathbb{R}^d$ and given with respect to the counting measure by
\begin{equation}
f\left(\x\mid\btheta,\G\right) = \frac{1}{Z\left(\btheta,\G\right)}\exp\left\lbrace- \pot_{\btheta,\G}\left(\x\right)\right\rbrace :=  \frac{1}{Z\left(\btheta,\G\right)}q(\x\mid\btheta,\G),
\label{eqn:gibbs}
\end{equation}
where $\pot_{\btheta,\G}$ denotes the energy function which can be written as a sum over the set $\mathcal{C}$ of all cliques of the graph, namely $\pot_{\btheta,\G}(\x) = \sum_{c\in\mathcal{C}}\pot_{\btheta,c}(\x)$ for all configuration $\x\in\X$. The inherent difficulty of all these models arises from the intractable normalizing constant, sometimes called the partition function, defined by
\[
Z(\btheta,\G)  = \sum_{\x\in \X} \exp\left\lbrace -\pot_{\btheta,\G}(\x) \right\rbrace.
\]
The latter is a summation over the numerous possible realisations of the random field $\Xv$, that is of combinatorial complexity and cannot be computed directly (except for small grids and small number of states $K$). For binary variables $X_i$, the number of possible configurations reaches $2^n$. 
 
\subsection{Autologistic model and related distributions}
\label{subsec:auto-models}

The formulation in terms of potential allows the local dependency of the Markov field to be specified and leads to a class of flexible parametric models for spatial data. In most cases, cliques of size one (singleton $c_0$) and two (doubletons $\C = \{c_1, c_2, c_3, c_4\}$) are assumed to be satisfactory to model the spatial dependency and potential functions related to larger cliques are set to zero. We present below popular examples of such pairwise models broadly used in the literature.

\paragraph*{Autologistic model--Ising model} The autologistic model first proposed by \cite{besag1972} is a pairwise-interaction Markov random field for binary (zero-one) spatial process. Denote $\btheta = (\alpha, \beta)$. The joint distribution is given by the following energy function
\begin{equation}
\pot_{\btheta,\G}(\x) = -\alpha\sum_{i=1}^n x_i - \beta\sum_{\ivj} x_i x_j,
\label{eqn:ising}
\end{equation}
where the above sum $\sum_{\ivj}$ ranges the set of edges of the graph $\G$.
The full-conditional probability thus writes
like a logistic regression where the explanatory variables are the neighbours and themselves observations. The parameter $\alpha$ controls the level of $0-1$ whereas the parameters $\{\beta\}$ model the dependency between two neighbouring sites $i$ and $j$. One usually prefers to consider variables taking values in $\{ -1,1\}$ instead of $\{0,1\}$ since it offers a more parsimonious parametrisation and avoids non-invariance issues when one switches states $0$ and $1$ as mentioned by \cite{pettitt2003}. 
A well known example is the general Ising model of ferromagnetism \citep{ising1925} that consists of discrete variables representing spins of atoms. The Gibbs distribution \eqref{eqn:ising} is sometimes referred to as the Boltzmann distribution in statistical physics. The potential on singletons describes local contributions from external fields to the total energy. Spins most likely line up in the same direction of $\alpha$, that is, in the positive, respectively negative, direction if $\alpha>0$, respectively $\alpha<0$. 
Putting differently $\alpha$ adjusts non-equal abundances of the two state values. The parameters $\{\beta\}$ represent the interaction strength between neighbours $i$ and $j$. When $\beta > 0$ the interaction is called ferromagnetic and adjacent spins tend to be aligned, that is neighbouring sites with same sign have higher probability. When $\beta < 0$ the interaction is called anti-ferromagnetic and adjacent spins tend to have opposite signs. 

\paragraph*{Potts model} 
The Potts model \citep{potts1952} originally appears in statistical mechanics to model interacting spins but has been used in other modelling area since then. It is a pairwise Markov random field that extends the Ising model to $K$ possibles states. The model sets a probability distribution on parametrized by $\btheta =(\alpha_0,\ldots,\alpha_{K-1}, \beta)$, namely
\begin{equation}
\pot_{\btheta,\G}(\x) = -\sum_{i = 1}^n \sum_{k=0}^{K-1}\alpha_{k}\ind\{x_i = k\} - \beta\sum_{\ivj}\ind\{x_i = x_j\},
\label{eqn:potts-pot}
\end{equation}
where $\ind\{A\}$ is the indicator function equal to 1 if $A$ is true and 0 otherwise. Note that a potential function can be defined up to an additive constant. To ensure that potential functions on singletons are uniquely determined, one usually imposes the constraint $\sum_{k=0}^{K-1}\alpha_k = 0$. The 2-states Potts model is equivalent to the Ising model up to a constant for interaction parameter $\beta$, that is $\beta_{\text{Potts}}=2\beta_{\text{Ising}}$.

\subsection{Phase transition}

%

One major peculiarity of Markov random field in the absence of an external field (\textit{i.e.}, $\alpha = 0$) is a symmetry breaking for large values of parameter $\beta$ due to a discontinuity of the partition function when the number of sites $n$ tends to infinity. When the parameter $\beta$ is zero, the random field is a system of independent uniform variables and all configurations are equally distributed. Increasing $\beta$ favours the variable $X_i$ to be equal to the dominant state among its neighbours and leads to patches of like-valued variables in the graph, such that once $\beta$ tends to infinity values $x_i$ are all equal. The distribution thus becomes multi-modal. In physics this is known as phase transition. This transition phenomenon has been widely study in both physics and probability, see for example \cite{georgii2011} for further details. The two dimensional Ising model is known to have a phase transition at a critical value $\beta_c$. \cite{onsager1944} obtained an exact value of $\beta_c$ for the Ising model on the first order square lattice, namely
\begin{equation*}
\beta_c = \frac{1}{2}\log\left\lbrace 1+\sqrt{2}\right\rbrace \approx 0.44.
\end{equation*}
The latter extends to a $K$-states Potts model on the first order lattice 
\begin{equation*}
\beta_c = \log\left\lbrace 1+\sqrt{K}\right\rbrace,
\end{equation*}
see for instance \cite{matveev1996} for specific results to Potts model on the square lattice and \cite{wu1982} for a broader overview. 

\begin{figure}[t]
\centering
\begin{minipage}[t]{7.cm}
\centering
\includegraphics[width=7.cm] {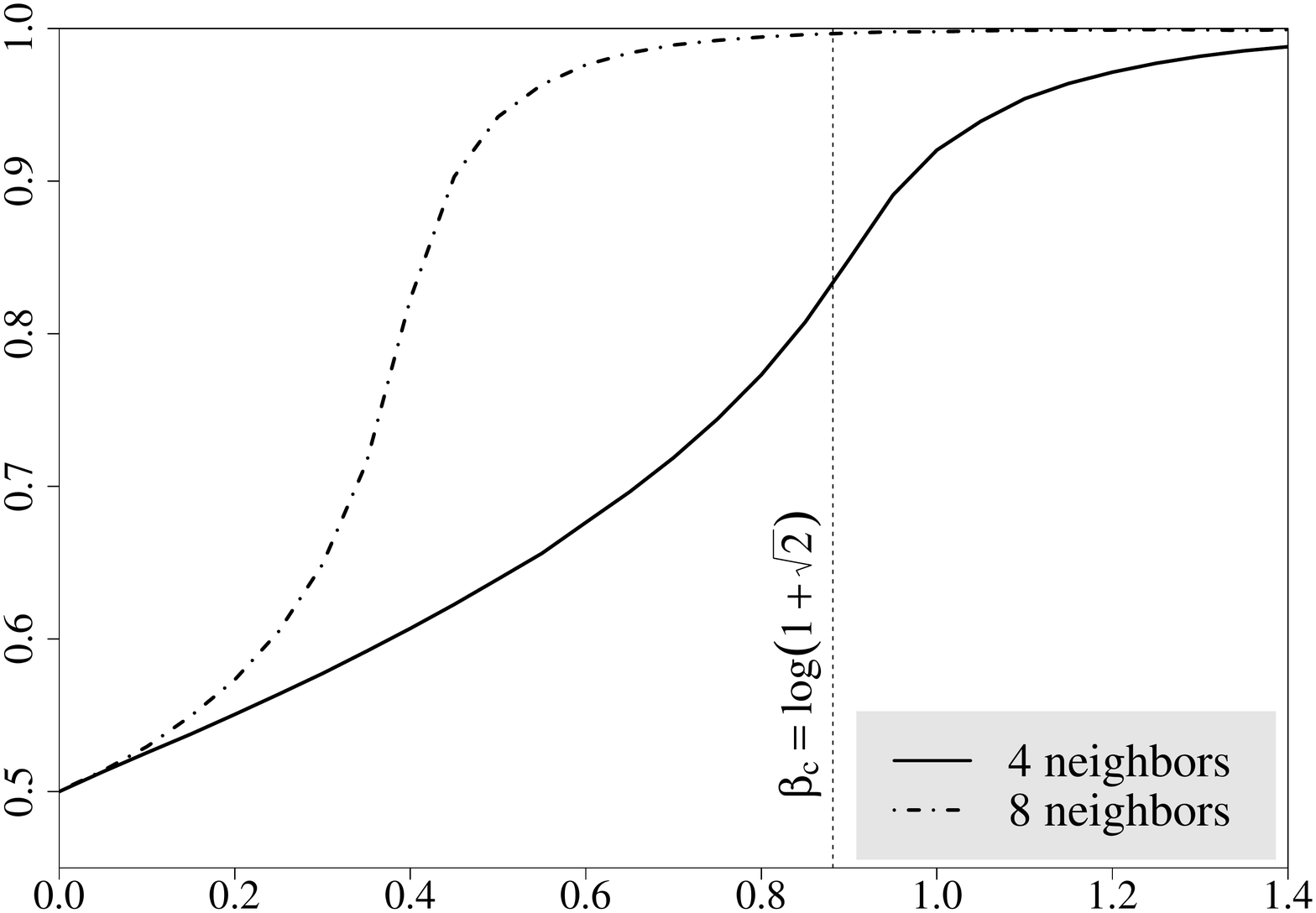}\\
(a)
\end{minipage}
\begin{minipage}[t]{7.cm}
\centering
\includegraphics[width=7.cm] {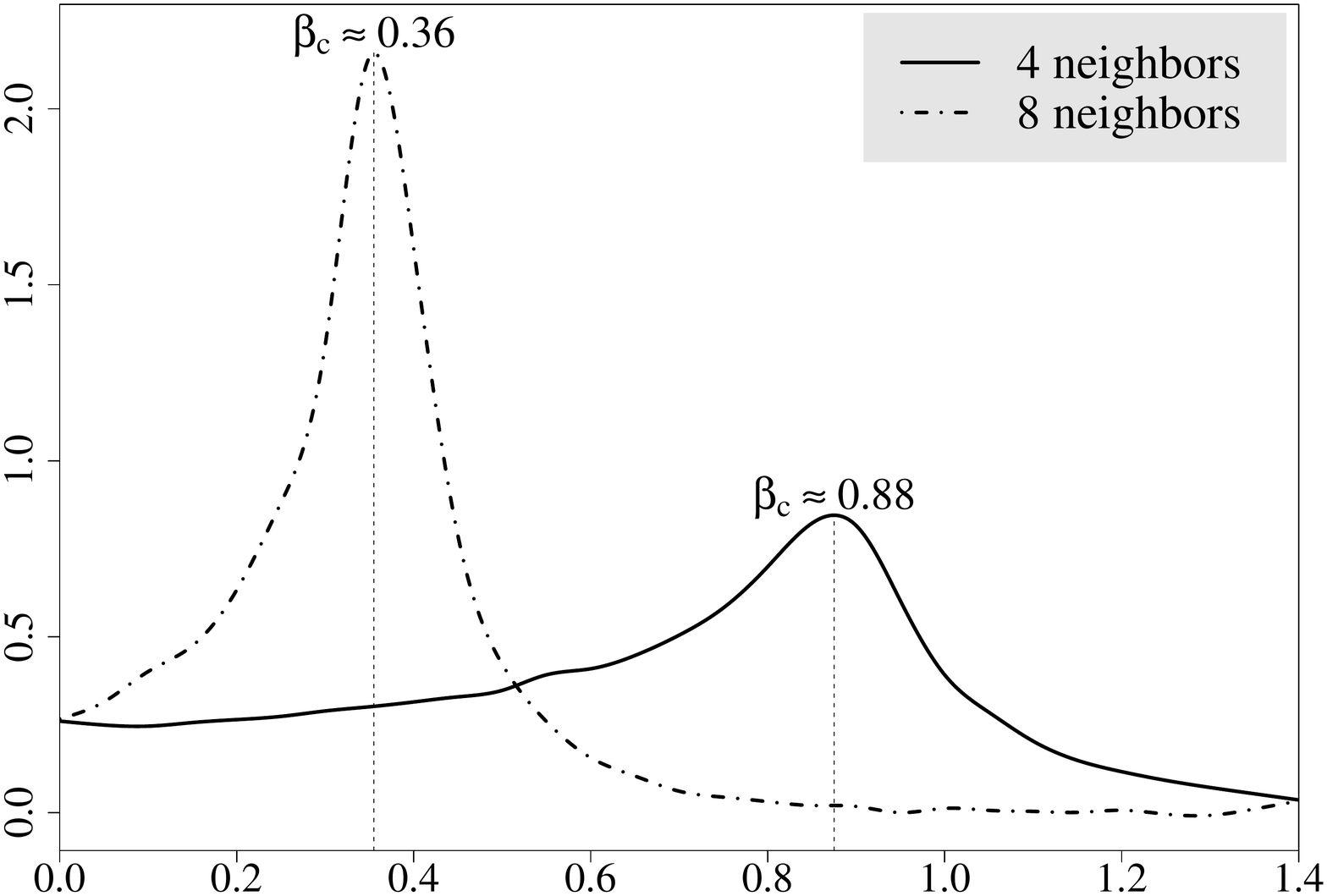}\\
(b)
\end{minipage}

\caption{\small Phase transition for a 2-states Potts model with respect to the first order and second order $100\times 100$ regular square lattices. (a) Average proportion of homogeneous pairs of neighbours. (b) Variance of the number of homogeneous pairs of neighbours.}
\label{fig:transition}
\end{figure}

The transition is more rapid than the number of neighbours increases. To illustrate this point, Figure \ref{fig:transition} gives the average proportion of homogeneous pairs of neighbours, and the corresponding variance, for 2-states Potts model on the first and second order lattices of size $100\times 100$. Indeed, phase transition corresponds to discontinuity at $\beta_c$ of $\beta\mapsto\lim_{n\rightarrow\infty}\frac{1}{n}\nabla\log Z\left(\beta,\G\right)$. Since $\hess\log Z(\beta,\G) = \var\left\lbrace\Ss(\Xv)\right\rbrace$, where $\Ss(\Xv)=\sum_{\ivj}\ind\{X_i=X_j\}$ is the number of homogeneous pairs of a Potts random field $\Xv$, the discontinuity condition can thus be written as 
\[
\lim_{\beta\rightarrow\beta_c}\lim_{n\rightarrow\infty} \var\left\lbrace\Ss(\Xv)\right\rbrace = \infty.
\]
Mention this is all theoritical asymptotic considerations and the discontinuity does not show itself on finite lattice realizations but the variance becomes increasingly sharper as the size grows.

\subsection{Hidden Gibbs random field}
\label{sec-hmrf}

Hidden Markov random fields has encountered a large interest over the past decade. It offer an appropriate representation for practical settings where the true state is unknown and observed indirectly through another field; this permits the modelling of noise that may happen upon many concrete situations: image analysis, \citep[\textit{e.g., }][]{besag1986, forbes2003, alfo2008, moores2014}, disease mapping \citep[\textit{e.g., }][]{green2002}, genetic analysis \citep{francois2006}, to name but a few. The unobserved data is modelled as a discrete Markov random field $\Xv$ associated to an energy function $\pot_{\btheta,\G}$, as defined in \eqref{eqn:gibbs}. Given the realization $\x$ of the latent, the observation $\y$ is a family of random variables indexed by the set of sites $\s$, and taking values in a set $\Y$, \textit{i.e.}, $\y=\left(y_i; {i\in\s}\right)$, and are commonly assumed as independent draws that form a noisy version of the hidden field. Consequently, we set the conditional distribution of
$\Yv$ knowing $\Xv=\x$, also called emission distribution, as the product 
\[
g_{\bphi}\left(\y\suchthat \x\right)=\prod_{i\in\s}
g_i\left(y_i\suchthat x_i,\bphi\right), 
\]
where $g_i$ is the marginal noise distribution parametrized by $\bphi$, that is given for any site $i$. Those marginal distributions are for instance discrete distributions \citep{everitt2012}, Gaussian \citep[\textit{e.g., }][]{besag1991, forbes2003, cucala2013} or Poisson distributions \citep[\textit{e.g., }][]{besag1991}. Model of noise that takes into account information of the nearest neighbours have also been explored \citep{besag1986}.

Assuming that all the marginal distributions $g_i$ are positive, one may write
the joint distribution of $(\Xv,\Yv)$, also called the complete likelihood, as
\begin{align*}
\pr\left(\x,\y\suchthat \btheta, \bphi, \G\right)  
& = \frac{1}{Z\left(\btheta,\G\right)}\exp\left\lbrace-\pot_{\btheta,\G}\left(\x\right) + \sum_{i\in\s}\log g_i\left(y_i\suchthat x_i,\bphi\right) \right\rbrace.&
\end{align*}
The conditional field $\Xv$ given $\Yv=\y$ is thus a Markov random field and the noise can be interpreted as a non homogeneous external potential on singleton which is a bond to the unobserved data.

\section{Statistical analysis issues}
\label{sec:issues}

The intractable normalising constant $Z(\btheta,\G)$ forms a central issue for both parameter and model selection problems as the likelihood is an integral part of the statistical procedure. Below, we introduce some of the classical problems of the literature

\paragraph*{Maximum likelihood estimator} Under the statistical model $f(\x\suchthat\btheta,\G)$, computing the maximum likelihood estimator, namely
\begin{equation}
\hat{\btheta}_{\text{MLE}} = \argmax_{\btheta} ~\log f\left(\x\suchthat\btheta,\G\right),
\label{eqn:mle}
\end{equation}
is challenging. Indeed, closed-form gradients are typically out of reach. Furthermore, one cannot rely on differentiation techniques such as automatic differentiation \cite[\eg][]{neidinger2010} since point-wise estimation is impossible due to the intractability issue.

\paragraph*{Computations of posterior distributions} Consider a Bayesian posterior distribution expressed as
\begin{equation}
\pi\left(\btheta\knowing \x\right) \propto f\left(\x\suchthat\btheta,\G\right) \pi(\btheta),
\label{eqn:posterior}
\end{equation}
where $f(\x\suchthat\btheta,\G)$ denoted the likelihood of the observed data $\x\in\X$ and $\pi(\btheta)$ denotes a prior density on the parameter space $\bTheta$ with 
respect to a reference measure (often the Lebesgue measure of the Euclidean space).
Here we are concerned with the situation where the un-normalised posterior distribution, the right-hand-side of (\ref{eqn:posterior}) is intractable. This complication results in what is often termed a doubly-intractable
posterior distribution, since the posterior distribution itself is normalised by the evidence (or marginal likelihood) which is typically also intractable. 

\paragraph*{Bayesian model choice} Model choice is a problem of probabilistic model comparison. Assume we are given a set $\M=\{m:~1,\ldots,M\}$ of stochastic models with respective parameter spaces $\bPsi_m$ embedded into Euclidean spaces of various dimensions. Bayesian approach to model selection considers the model itself as an unknown parameter of interest. 
The joint Bayesian distribution sets: a prior $\pi$ on the model space, a prior density $\pi_m$ on each parameter space $\bPsi_m$ with respect to a reference measure (often the Lebesgue measure of the Euclidean space), the likelihood $f_m$ of the data $\y$ within each model. On the extended parameter space $\bPsi=\bigcup_{m=1}^M\{m\}\times\bpsi_m$, the Bayesian analysis targets posterior model probabilities, that is the marginal in $\M$ of the posterior distribution for $(m, \bpsi_1,\ldots,\bpsi_M)$ given $\Yv = \y$, 
\[
\pi(m\mid\y)=\frac{ e(\y\mid m) \pi(m)}{\sum_{m'=1}^{M} e(\y\mid m')\pi(m')},
\]
where $e(\y\mid m)$ denotes the evidence (or integrated likelihood) of model $m$ defined as
\begin{equation}
e(\y\mid m)=\int_{\bPsi_m} f_m(\y\mid\bpsi_m)\pi_m(\bpsi_m)\mathrm{d}\bpsi_m.
\label{eqn:evidence}
\end{equation}
When the goal of the Bayesian analysis is the selection of the model that best fits the observed data $\y$, it is performed through the maximum \textit{a posteriori} (MAP) defined by 
\begin{equation}
\widehat{m}_\text{MAP}(\y)=\argmax_m \pi(m\mid\y).
\label{eq:bayes-classifier}
\end{equation}
The standard approach to compare one model against another is based on the Bayes factor \citep{kass-raftery1995} that involves the ratio of the evidence \eqref{eqn:evidence} of each model. Under the assumption of model being equally likely \textit{a priori}, the MAP rule \eqref{eq:bayes-classifier} is equivalent to choose the model with the largest evidence \eqref{eqn:evidence} and in place of a fully Bayesian approach, model choice criterion can be used. However, the evidence can usually not be computed with standard procedure because of a high-dimensional integral. Thus much of the research in model selection area focuses on evaluating it by numerical methods. 

Selecting the model among a collection of Markov random fields is a daunting task as none of \eqref{eqn:evidence} and \eqref{eq:bayes-classifier} are analytically available. The model selection problem for hidden Markov random fields is even more complicated and could be termed as a triply-intractable problem. Indeed in addition to the integral on $\bPsi_m$ which is typically intractable, the stochastic model for $\Yv$ is based on the latent process $\Xv$ in $\X$, that is
\begin{equation}
f_m(\y\mid\bpsi_m = (\btheta_m,\bphi_m)) = \int_{\X} g_{\bphi_m}(\y\mid\x) f(\x\suchthat\btheta_m,\G_m)\mu(\mathrm{d}\x),
\label{eqn:int-like}
\end{equation}
with $\mu$ the counting measure (discrete case). Both the integral and the Gibbs distribution are intractable and consequently so is the posterior distribution.

\section{Recursive algorithm for small Markov random field}
\label{sec:recursive}

When the Markov random field is defined on a small enough lattices, it is possible to answer the difficulty of computing the normalising constant by relying on generalised recursions for general factorisable models \cite{reeves2004}. This method is based on an algebraic simplification due to the reduction in dependence arising from the Markov property. It applies to unnormalized likelihoods that can be expressed as a product of factors, each of which is dependent on only a subset of the lattice sites. More specifically the unnormalized version of a Gibbs distribution can be write as
\begin{equation*}
q\left(\x\suchthat\btheta, \G\right) = \prod_{i=1}^{n-r} q_i\left(\x_{i:i+r}\suchthat\btheta, \G\right),
\end{equation*}
where each factor $q_i$ depends on a subset $\x_{i:r} = \{x_i,\dots,x_{i+r}\}$ of $\x$, where $r$ is defined to be the \emph{lag} of the model. 

As a result of this factorisation, the summation for the normalizing constant can be represented as
\begin{equation*}
Z\left(\btheta,\G\right) = \sum_{\x_{n-r:n}} q_{n-r}\left(\x_{n-r:n}\suchthat\btheta, \G\right)  \dots  \sum_{\x_{1:1+r}}q_1\left(\x_{1:1+r}\suchthat\btheta, \G\right).
\label{eqn:z_theta}
\end{equation*}
The latter can be computed much more efficiently than the straightforward summation over the $K^n$ possible lattice realisations using the following steps
\begin{align*}
& Z_1\left(\x_{2:1+r}\right) = \sum_{x_1} q_1\left(\x_{1:1+r}\right), \\
& Z_i\left(\x_{i+1:i+r}\right) = \sum_{x_i} q_i\left(\x_{i:i+r}\right)Z_{i-1}\left(\x_{i:i+r-1}\right), \text{ for all } i \in\{2,\ldots,n-r\}, \\
& Z\left(\psi, \G\right) = \sum_{\x_{n-r+1:n}} Z_{n-r}\left(\x_{n-r+1:n}\right).
\end{align*}

The complexity of the troublesome summation is significantly cut down since the forward algorithm solely relies on $K^r$ possible configurations. Consider a rectangular lattice $h\times w = n$, where $h$ stands for the height and $w$ for the width of the lattice, with a first order neighbourhood system $\G_4$ (see Figure \ref{fig:neigh}.(a)). The minimum \emph{lag} representation for a pairwise model define on such a lattice occurs for $r$ given by the smaller of the number of rows or columns in the lattice. Without the loss of generality, assume $h \leq w$ and lattice points are ordered from top to bottom in each column and columns from left to right. The complexity in time of the algorithm is then exponential in the number of rows and linear in the number of column. The algorithm of \cite{reeves2004} was extended in \cite{friel2007} to also allow exact draws from $\pi(\x\mid\psi, \G)$ for small enough lattices. The reader can find below an example of implementation for the general Potts model.


Following the work of \cite{friel2007}, the R-package GiRaF available on CRAN proposes amongst other tools an ingenious implementation of those recursions for the autologistic, the Ising and the Potts model. Indeed, a naive implementation of the aforementioned recursions for such models can substantially increase the cost in time and memory allocation. For instance, the unnormalized Potts distribution associated to energy function \eqref{eqn:potts-pot} writes as \[
q\left(\x\suchthat\psi, \G_4\right) = \prod_{i=1}^{n-h} q_i\left(\x_{i:i+h}\suchthat\psi, \G_4\right),
\] where
\begin{itemize}
\item for all lattice point $i$ except the ones on the last row or last column
\begin{multline}
q_i\left(\x_{i:i+h}\suchthat\psi, \G_4\right) = \exp\Bigg(\sum_{k=0}^{K-1}\alpha_k \ind\{x_i=k\} \\ + \beta_{0} \ind\{x_i=x_{i+1}\}
+ \beta_{1} \ind\{x_i=x_{i+h}\}\Bigg).
\label{eqn:factor-rec}
\end{multline}
\item When lattice point $i$ is on the last row $x_{i+1}$ drops out ot \eqref{eqn:factor-rec}, that is
\begin{equation}
q_i\left(\x_{i:i+h}\suchthat\psi, \G_4\right) = \exp\left(\sum_{k=0}^{K-1}\alpha_k \ind\{x_i=k\}
+ \beta_{1} \ind\{x_i=x_{i+h}\}\right).
\label{eqn:factor-rec2}
\end{equation}
\item The last factor takes into account all potentials within the last column
\begin{multline*}
q_{n-h}\left(\x_{n-h:n}\suchthat\psi, \G_4\right) = \exp\Bigg(\sum_{i=n-h}^n\sum_{k=0}^{K-1}\alpha_k \ind\{x_i=k\} \\
+ \beta_{1} \ind\{x_{n-h}=x_{n}\} + \beta_0\sum_{i=n-h+1}^n \ind\{x_i=x_{i+1}\}\Bigg).
\end{multline*}
\end{itemize}
One shall remark that for a homogeneous random field (\ie parameters are independent of the location of the sites), factors \eqref{eqn:factor-rec} and \eqref{eqn:factor-rec2} only depend on the value of the random variables $\Xv_{i:i+h}$ but not on the actual position of the sites. Hence the number of factors to be computed is $2K^h$ instead of $h(w-1)K^h$. In term of implementation that also means factors can be computed for the different possible configurations once upstream the recursion. Furthermore with a first order neighbourhood, factor at a site merely involves its neighbour below and on its right, thereby reducing the number of possible factor to $K^3+K^2$. 
Finally, mention it is straightforward to extend this algorithm to hidden Markov random field since as already mention in Section \ref{sec-hmrf} the noise corresponds to a non homogeneous potential on singleton and hence the model still writes as a general factorisable model. 

\section{Pseudo-model and variational approaches}
\label{sec:pseudo-model}

A point of view to overcome model’s bottleneck is to replace the true model with another pseudo-model selected among a collection of much simpler probability distribution, like for variational Bayes \citep{jaakkola2000}, or with an easily-normalised full conditional distributions \citep{lindsay1988}. Both options have been explored in the literature and we discuss some of the solutions below. 

\subsection{Composite likelihood}

A composite likelihood \citep{lindsay1988} approximates the joint distribution as the product of easy normalised full-conditional distributions
\begin{equation}
 f_{\text{CL}}\left(\x\suchthat\btheta, \G\right) = \prod_{i=1}^C f\left(\x_{A(i)}\suchthat \x_{B(i)},\btheta,\G\right)^{w_i},
\label{eqn:approx-lindsay}
\end{equation}
where $\big\lbrace A(i)\big\rbrace_{i=1}^C \subseteq \mathcal{P}(\s)$ and $\big\{B(i)\big\}_{i=1}^C\subseteq \mathcal{P}(\s)$ denote sets of subset of $\s$. It has encountered considerable interests in the statistics literature and the reader may refer to \cite{varin2011} for a comprehensive overview. One of the earliest approach using composite likelihood is the pseudolikelihood \citep{besag1975} 
which approximates the joint distribution of $\x$ as the product of full-conditional distributions for each site $i$,
\begin{equation}
f_{\text{pseudo}}\left(\x\suchthat \btheta, \G\right)= \prod_{i=1}^n f\left(x_i\suchthat \x_{-i},\btheta,\G\right).
\label{eqn:pseudo}
\end{equation}
The pseudolikelihood \eqref{eqn:pseudo} is not a genuine probability distribution, except if the random variables $X_i$ are independent. The Markov property ensures that each term in the product only involves nearest neighbours, and so the normalising constant of each full-conditional is straightforward to compute. It is straightforward to prove that a unique maximum exists and it is easy to compute.

\cite{geman1986} demonstrate the consistency of the maximum pseudolikelihood estimator 
\[
\hat{\btheta}_{\text{MPLE}} = \argmax_{\btheta}~ \log f_{\text{pseudo}}\left(\x\suchthat \btheta, \G\right).
\]
when the lattice size tends to infinity for discrete Markov random field
but maximum pseudolikelihood estimator has generally larger asymptotic variance than maximum likelihood estimator and does not achieve the Cramer-Rao lower bound \citep{lindsay1988}. Practically, this approximation has been shown to lead to unreliable estimates of $\btheta$ \citep[\textit{e.g.}, ][]{ryden1998, friel2004, cucala2009, friel2009}. 

\subsection{Variational approaches and parameter estimation}
\label{sec:variational}

Variational methods refer to a class of deterministic approaches. They consist in introducing a variational function as an approximation to the likelihood in order to solve a simplified version of an optimization problem. It has long-standing antecedents in statistical mechanics when one aims at predicting the response to the system to a change in the Hamiltonian. One important technique is based on a variational approach as the minimizer of the free energy, sometimes referred to as variational or Gibbs free energy and defined with the Kullback--Leibler divergence between a probability distribution $\pr$ and the target distribution $f(\cdot\suchthat\btheta,\G)$ as
\begin{equation}
F(\pr)=-\log Z\left(\btheta, \G\right) + \KL\left(\pr\Vert f(\cdot\suchthat\btheta,\G)\right).
\label{eqn:free-energy}
\end{equation}
Although the Kullback--Leibler divergence is not a true metric, it has the non-negative property with divergence zero if and only if distributions are equal almost everywhere. The free energy has then an optimal lower bound achieved for $\pr = f(\cdot\suchthat\btheta,\G)$. Minimizing the free energy with respect to the set of probability distribution on $\X$ allows to recover the Gibbs distribution but presents the same computational intractability. A solution is to minimize the Kullback--Leibler divergence over a restricted class of tractable probability distribution on $\X$. This is the basis of mean field approaches that aim at minimizing the Kullback--Leibler divergence over the set of probability functions that factorize on sites of the lattice.
namely for all $\x$ in $\X=\prod_{i\in\s}\X_i$,
\[
\pr(\x)=\prod_{i\in\s}\pr_i(x_i), \text{ where } \pr_i\in\mathcal{M}_{1}^{+}(\X_i)\text{ and }\pr\in\mathcal{M}_1^{+}(\X). 
\]
The minimization of \eqref{eqn:free-energy} over this set leads to fixed point equations for each marginal of $\pr$ and the optimal solution is the so called mean field approximation \citep[see for example][]{jordan1999}. 

\subsubsection*{Variational EM algorithm}

In what follows we focus on the issue of estimating the parameters of a hidden Markov random fields. Let consider $\y$ a noisy version of a Markov random field $\x$. One can write the log-likelihood as follows
\begin{equation}
\log f\left(\y\suchthat\bpsi\right) = \underbrace{\sum_{\x\in\X}\log\left\lbrace\frac{f\left(\x,\y\suchthat\bpsi,\G\right)}{\pr(\x)}\right\rbrace \pr(\x)}_{=F(\pr,\bpsi)}
+ \underbrace{\sum_{\x\in\X}\log\left\lbrace\frac{\pr(\x)}{f\left(\x\suchthat\y,\bpsi,\G\right)}\right\rbrace \pr(\x)}_{=\KL\left(\pr\Vert~f(\cdot\mid\y,\bpsi,\G)\right)}.
\label{eqn:variational}
\end{equation}
Solutions based on variational approaches focus on minimising the KL-term or equivalently maximising the  $F$. For instance, this relaxation of the original issue has shown good performances for approximating the maximum likelihood estimate \citep{celeux2003}, as well as for Bayesian inference on hidden Markov random fields \citep{mcgrory2009}. We discuss it below.

\cite{celeux2003} explore the opportunity to use the Expectation-Maximization (EM) algorithm \citep{dempster1977}. To address the problem of finding the maximum likelihood estimator for the statistical model $f(\y\mid\bpsi)$, the EM  
algorithm iterates between two steps:
\begin{enumerate}
\item \textbf{E step:} one computes the conditional expectation of the log-joint distribution with respect to the distribution of $\Xv$ given $\Yv =\y$ at the current parameter value $\bpsi^{(t)}$
\begin{align*}
 Q(\bpsi\suchthat\bpsi^{(t)}) & = \esp_{\bpsi^{(t)}}\left\lbrace\log f\left(\Xv,\y\suchthat\bpsi,\G\right)\suchthat \Yv=y\right\rbrace  \\
& = \sum_{\x\in\X} f(\x\suchthat\y,\bpsi^{(t)},\G)\log f\left(\x,\y\suchthat\bpsi,\G\right).
\end{align*}
\item \textbf{M step:} one maximises $Q$ with respect to $\bpsi$,
\[
\bpsi^{(t+1)} = \argmax_{\bpsi} Q\left(\bpsi\suchthat\bpsi^{(t)}\right).
\]
\end{enumerate}
We refer the reader to \cite{wu1983} for convergence results. The EM scheme cannot be applied directly to hidden Markov random fields as it yields analytically intractable updates. The function $Q$ can be written as
\begin{align*}
Q(\bpsi\suchthat \bpsi^{(t)}) & = \esp_{\bpsi^{(t)}}\left\lbrace\log g_{\bphi}\left(\y\suchthat \Xv\right)\suchthat \Yv = \y,\right\rbrace
+ \esp_{\bpsi^{(t)}}\left\lbrace\log f\left(\Xv\suchthat \btheta,\G\right)\suchthat \Yv = \y \right\rbrace.
\end{align*}
and hence requires evaluating $f(\cdot\mid\btheta,\G)$ and $f(\cdot\mid\y,\bpsi,\G)$ (for computing $\esp_{\bpsi}$) which are both unavailable. Many stochastic or deterministic schemes have been proposed to handle intractability in the EM steps, such as the Gibbsian-EM \citep{chalmond1989}, the Monte-Carlo EM \citep{wei1990} or the Restoration-Maximization algorithm \citep{qian1991}.
The aim of the variational EM (VEM) is to maximize the function $F$ instead of $Q$ in order to get a tractable version of the EM algorithm. This shift in the formulation leads to an alternating optimization procedure which can be described as follows: let denote $\mathcal{D}$ a set of probability distributions on the latent space , given a current value $(\pr^{(t)}, \theta^{(t)})$ in $\D\times\Theta$, updates with
\begin{align}
\label{eqn:vem-q}
\pr^{(t+1)} & = \argmax_{\pr\in\D} F(\pr,\bpsi^{(t)}) = \argmin_{\pr\in\D} \KL\left(\pr\Vert~f(\cdot\mid\y,\bpsi^{(t)},\G)\right), \\
\bpsi^{(t+1)} & = \argmax_{\bpsi} F\left(\pr^{(t+1)},\bpsi\right) = \argmax_{\theta} \sum_{\x\in\X}\pr^{(t+1)}(\x)\log\pi\left(\x,\y\suchthat\bpsi,\G\right). 
\label{eqn:vem-theta}
\end{align}
Optimising $F$ over the class of independent probability distributions $\pr$ that factorize on sites leads to mean field approximation. Generalizing an idea originally introduced by \cite{zhang1992}, \cite{celeux2003} have designed a class of VEM-like algorithm that uses mean field-like approximations for both $f(\cdot\mid\y,\bpsi,\G)$ and $f(\cdot\mid\btheta,\G)$. To put it in simple terms mean field-like approximations refer to distributions for which neighbours of site $i$ are set to constants. Given a configuration $\tilde{\x}$ in $\X$, the Gibbs distribution $f(\cdot\mid\btheta,\G)$ is replaced by
\[
\pr^{\mfl}_{\tilde{\x}}\left(\x\suchthat \btheta, \G\right) = \prod_{i\in\s}\pr\left(x_{i};\tilde{\x}_{\N(i)},\btheta,\G\right).
\]
The main difference with the pseudolikelihood \eqref{eqn:pseudo} is that neighbours are not random anymore and setting them to constant values leads to a system of independent variables. From this approximation, the EM path is set up with the corresponding joint distribution approximation
\[
\pr^{\mfl}\left(\x,\y\suchthat \bpsi,\G\right) = \prod_{i\in\s} g_{\bphi}\left(y_i\suchthat x_i\right) \pr\left(x_{i};\tilde{\x}_{\N(i)},\btheta,\G\right).
\]
Note that this general procedure corresponds to the so-called point-pseudo-likelihood EM algorithm proposed by \cite{qian1991}. 
The flexibility of the approach proposed by \cite{celeux2003} lies in the choice of the configuration $\tilde{\x}$ that is not necessarily a valid configuration for the model. We refer the reader to \cite{celeux2003} for further details. When the neighbours $\Xv_{\N(i)}$ are fixed to their mean value, or more precisely $\tilde{\x}$ is set to the mean field estimate of the complete conditional distribution $f(\x\mid\y,\bpsi,\G)$, this results in the Mean Field algorithm of \cite{zhang1992}. In practice, \cite{celeux2003} obtain better performances with their so-called Simulated Field algorithm (see Algorithm \ref{algo:simulated-field}). In this stochastic version of the EM-like procedure, $\tilde{\x}$ is a realization drawn from the conditional distribution $f(\cdot\mid\y,\bpsi^{(t)},\G)$ for the current value of the parameter $\bpsi^{(t)}$. The latter is preferred to other methods when dealing with maximum-likelihood estimation for hidden Markov random field. This extension of VEM algorithms suffers from a lack of theoretical support due to the propagation of the approximation to the Gibbs distribution $f(\cdot\mid\btheta\G)$. One might advocate in favour of the Monte-Carlo VEM algorithm of \cite{forbes2007} for which convergence results are available. However the Simulated Field algorithm provides better results for the estimation of the spatial parameter, as illustrated in \cite{forbes2007}.

\begin{algorithm2e}[!t]
  \caption{Simulated Field algorithm}
  \label{algo:simulated-field}
  \KwIn{an observation $\y$, a number of iterations $T$}
  \KwOut{an estimate of the likelihood maximum estimator $\hat{\psi}_{\mle}$ }
  \medskip
  
  {\bf Initialization:} start from an initial guess $\bpsi^{(0)}=\left(\btheta^{(0)},\bphi^{(0)}\right)$\;
  \For{$t\leftarrow 1$ \KwTo $T$}{
  {\bf neighbourhood restoration:} draw $\tilde{\x}^{(t)}$ from $\pi\left(\cdot\suchthat\y,\bpsi^{(t-1)},\G\right)$\;
  {\bf E-step:} compute 
  \begin{align*}
\widehat{Q_1}(\bphi) 
 := \sum_{i\in\s}\sum_{x_i} \pr\big(x_i;\tilde{\x}^{(t)}_{\N(i)}, y_i, & \bpsi^{(t-1)},\G\big)\log g_{\bphi}\left(y_i\suchthat x_i\right); \\
\widehat{Q_2}(\btheta)
 := \sum_{i\in\s}\sum_{x_i} \pr\big(x_i; \tilde{\x}^{(t)}_{\N(i)}, y_i, & \bpsi^{(t-1)},\G\big) \\
& \log\pr\left(x_{i};\tilde{\x}^{(t)}_{\N(i)},\btheta,\G\right);
\end{align*}

  {\bf M-step:} set $\bpsi^{(t)}=\left(\btheta^{(t)},\bphi^{(t)}\right)$ where 
  \[\btheta^{(t)}=\argmax_{\btheta} \widehat{Q_1}(\btheta) \text{ and }\bpsi^{(t)}=\argmax_{\bpsi} \widehat{Q_2}(\bpsi);\]
  }
  \textbf{return} $\bpsi^{(T)}=\left(\btheta^{(T)},\bphi^{(T)}\right)$ 
\end{algorithm2e}

\subsection{Approximating model choice criteria}
\label{subsec:bic-mf}

Various approximations have been proposed to overcome intractability of \eqref{eqn:evidence} but a commonly used one, if only for its simplicity, is the Bayesian Information Criterion (BIC) that is an asymptotic estimate of the evidence based on the Laplace method for integrals \citep{schwarz1978, kass-raftery1995}. The criterion is a simple penalized function of the maximized log-likelihood 
\begin{equation}
\BIC(m) = -2\log f_m\left(\y\suchthat \hat{\bpsi}_{\mle}\right) + d_m\log(n) \approx -2\log e(\y\mid m),
\label{eqn:bic}
\end{equation}
where $d_m$ is the number of free parameters of model $m$ (usually the dimension of $\bPsi_m$) and $n=\vert\s\vert$ is the number of sites. The $d_m \log(n)$ term corresponds to a penalty term which increases with the complexity of the model and the model with the highest posterior probability is the one that minimizes BIC. The criterion is closely related to the Akaike Information Criterion \citep[AIC,][]{akaike1973} that solely differs in the penalization term. We refer the reader to \cite{kass-raftery1995} and the references therein for a more detailed discussion on AIC and for instance to \cite{burnham2002} for comparison between AIC and BIC.

BIC can be defined beside the special case of independent random variables. In the latter case the number of free parameter is, in general, not equal to the dimension of the parameter space as for the independent case. The consistency of BIC has been proven in various situations such as independent and identically distributed processes from the exponential families \citep{haughton1988}, mixture models \citep{keribin2000} or Markov chains \citep{csiszar2000, gassiat2002}. When dealing with observed Markov random fields, aside from the problem of intractable likelihoods the number of free parameters in the penalty term has no simple formula. In the context of selecting a neighbourhood system, \cite{csiszar2006} proposed to replace the likelihood by the pseudolikelihood \eqref{eqn:pseudo} and modify the penalty term as the number of all possible configurations for the neighbouring sites. The resulting criterion is shown to be consistent as regards this model choice. Up to our knowledge such a result has not been yet derived for hidden Markov random fields. 

As already mention, in the context of Markov random fields, difficulties are of two kind. Neither the maximum likelihood estimate $\hat{\btheta}_{\mle}$ nor the likelihood $f_m$ are available. Recall that in the hidden case $f_m$ requires to integrate a Gibbs distribution over the latent space configuration. As regards the simplest case of observed Markov random field solutions have been brought by penalized pseudolikelihood \citep{ji1996} or MCMC approximation of BIC \citep{seymour1996}. Over the past decade, only few works have addressed the model choice issue for hidden Markov random field from that BIC perspective. Below, we describe solutions based on pseudo-models but other attempts based on simulations techniques have been investigated \citep{newton1994}. 

The central question is the evaluation of the integrated likelihood \eqref{eqn:int-like}. A convenient way to circumvent the issues of computing BIC is to replace the Gibbs distribution by tractable surrogates. Consider a partition of $\s$ into subsets of neighbouring sites, namely
\[
\s = \bigsqcup_{\ell=1}^C A(\ell),
\] 
and denote by $\tilde{D}$ the class of independent probability distributions $\pr$ that factorize with respect to this partition, that is if $\X_{A(\ell)} $ stands for the configuration space of the subset $A(\ell)$, for all $\x$ in $\X$
\[
\pr(\x)=\prod_{\ell = 1}^C \pr_\ell\left(x_{A(\ell)}\right), \text{ where } \pr_{\ell}\in\mathcal{M}_{1}^{+}\left(\X_{A(\ell)}\right)
\text{ and }\pr\in\mathcal{M}_1^{+}(\X). 
\]
The most straightforward approach is to look for mean-field like approximations which take the following form
\[
\pr_{\tilde{\x}}(\x\mid\btheta,\G) = \prod_{i=1}^n\pr(x_{i} ; \tilde{\x}_{\N(i)},\btheta,\G) \approx f(\x\suchthat\btheta,\G).
\]
Put in other words, the latter is a surrogate in the class of independent probability distributions that factorize with respect to the nodes of $\G$ and for which neighbourhood of each site has been set to a constant field $\tilde{\x}$. One shall remark we recover the mean-field approximations presented in Section \ref{sec:variational} when $\tilde{\x}$ is set to the mean-field realisation. The integrated likelihood corresponding to $\pr_{\tilde{\x}}$ is of the form
\[
\pr_m^{\mfl}\left(\y\suchthat\bpsi_m = (\btheta_m,\bphi_m)\right) = \prod_{i\in\s}\sum_{x_i}g_{\bphi_m}\left(y_i\suchthat x_i\right)\pr(x_{i} ; \tilde{\x}_{\N(i)},\btheta_m,\G).
\]
This results in the following approximation of BIC
\begin{equation}
\BIC^{\mfl}(m) = -2\log \pr_m^{\mfl}\left(\y\suchthat \hat{\bpsi}_{\mle}\right) + d_m\log(n).
\label{eqn:bicp}
\end{equation}
This approach includes the Pseudolikelihood Information Criterion (PLIC) of \cite{stanford2002} as well as the mean field-like approximations of BIC proposed by \cite{forbes2003}. The main difference between these criterion lies in the estimation $\bpsi_{\mle}$ and the choice of $\tilde{x}$.
The idea of \cite{stanford2002} is to consider as $\tilde{\x}$ a configuration close to the Iterated Conditional Modes \citep[ICM,][]{besag1986} estimate of $\x$. In its unsupervised version ICM alternates between a restoration step of the latent states and an estimation step of the parameter $\bpsi$. PLIC is the approximation of BIC based on the output of ICM algorithm $\left(\x^{\text{ICM}}, \bpsi^{\text{ICM}}\right)$. 
The solution proposed by \cite{forbes2003} is to use for $(\tilde{\x}, \hat{\bpsi}_{\mle})$ the output of the VEM-like algorithm based on the mean-field like approximations of \citep{celeux2003}. More precisely, as regards neighbourhood restoration step, they advocate in favor of the simulated field algorithm (see Algorithm \ref{algo:simulated-field}).

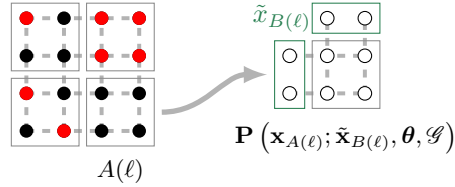
\begin{figure}[t]
\centering
\begin{minipage}[t]{7cm}
\centering
\begin{tikzpicture}

\foreach \y in {0,...,3}
	\draw[dashed, line width=1.5pt, gray!60] (0,0.5*\y) to[out=0,in=180] (1.5,0.5*\y);	
	
\foreach \y in {0,...,3}
	\draw[dashed, line width=1.5pt, gray!60] (0.5*\y, 0) to[out=90,in=-90] (0.5*\y, 1.5);	

\foreach \x in {0,...,3} 
	\foreach \y in {0,...,3}
   		\fill[black] (0.5*\x,0.5*\y) circle (0.9mm); 
   		
\fill[red] (0,0.5) circle (0.9mm);
\fill[red] (0.5,0) circle (0.9mm); 
\fill[red] (0,1.5) circle (0.9mm); 
\fill[red] (1,1) circle (0.9mm); 
\fill[red] (1.5,1) circle (0.9mm); 
\fill[red] (1.5,1.5) circle (0.9mm); 
\fill[red] (1,1.5) circle (0.9mm); 

\draw [gray!90] (-0.2,-0.2) rectangle (0.7,0.7);
\draw [gray!90] (0.8,-0.2) rectangle (1.7,0.7);
\draw [gray!90] (-0.2,0.8) rectangle (0.7,1.7);
\draw [gray!90] (0.8,0.8) rectangle (1.7,1.7);

\node[font = \small, below] at (1.25,-0.25) {$A(\ell)$};
\draw[-latex, line width=2pt, gray!60] (1.8, 0.25) to[out=0,in=180] (3.2,0.75);
\draw[dashed, line width=1.5pt, gray!60] (3.5,0.5) to[out=0,in=180] (4.5,0.5);	
\draw[dashed, line width=1.5pt, gray!60] (3.5,1) to[out=0,in=180] (4.5,1);
	
\draw[dashed, line width=1.5pt, gray!60] (4, 0.5) to[out=90,in=-90] (4, 1.5);	
\draw[dashed, line width=1.5pt, gray!60] (4.5, 0.5) to[out=90,in=-90] (4.5, 1.5);	

\draw[fill=white] (3.5,0.5) circle (0.9mm);
\draw[fill=white] (3.5,1) circle (0.9mm);
\draw [mygreen] (3.3,0.3) rectangle (3.7,1.2);
\draw[fill=white] (4,1.5) circle (0.9mm);
\draw[fill=white] (4.5,1.5) circle (0.9mm);
\draw [mygreen] (3.8,1.3) rectangle (4.7,1.7);
\node at (3.4,1.5) {\textcolor{mygreen}{$\tilde{x}_{B(\ell)}$}};

\node[font = \small, below] at (4.25, 0.25) {\small $\pr\left(\x_{A(\ell)}; \tilde{\x}_{B(\ell)},\btheta,\G\right)$};

\draw[fill=white] (4, 0.5) circle (0.9mm);
\draw[fill=white] (4.5, 0.5) circle (0.9mm);
\draw[fill=white] (4, 1) circle (0.9mm);
\draw[fill=white] (4.5, 1) circle (0.9mm);
\draw [gray!90] (3.8,0.3) rectangle (4.7,1.2);

\end{tikzpicture}

\end{minipage}
\caption{\small Factorisation of the Gibbs distribution approximation over a set of contiguous block $A(\ell)$ with border $B(\ell)$ set to a constant field $\tilde{\x}$.}
\label{fig:blic}
\end{figure}

\cite{stoehr2016} have extended previous approaches to tractable approximations that factorize over larger sets of nodes, namely blocks of a rectangular lattice, by taking advantage of the general recursion implemented in the R-package GiRaF (see Section \ref{sec:recursive}). They suggest to chose surrogates of the form 
\begin{equation}
\pr\left(\x\,;\,\tilde{\x}, A(1),\ldots,A(C),\btheta\right) = \prod_{\ell = 1}^C \pr\left(\x_{A(\ell)};\Xv_{B(\ell)} = \tilde{\x}_{B(\ell)}, \btheta,\G\right),
\label{eqn:product-distrib}
\end{equation}
where $B(\ell)$ is the border of $A(\ell)$, see Figure \ref{fig:blic}, or the empty set. This leads to their so called Block Likelihood Information Criterion (BLIC) which approximates BIC as a summation of tractable normalising constant, namely
\begin{equation}
\mathrm{BLIC}\left(m\right) = -2\sum_{\ell=1}^C  \frac{\log Z\left(\hat{\bpsi}_{\mle},\G,\y_{A(\ell)},\tilde{\x}_{B(\ell)}\right)}{\log Z\left(\hat{\btheta}_{\mle},\G, \tilde{\x}_{B(\ell)}\right)} + d_m\log(\vert\s\vert),
\label{eqn:bclic}
\end{equation}
where $Z\left(\btheta,\G, \tilde{\x}_{B(\ell)}\right)$ is the partition function of block $A(\ell)$ with fixed border $\tilde{\x}_{B(\ell)}$ and $Z\left(\theta,\G,\y_{A(\ell)},\tilde{\x}_{B(\ell)}\right)$ is the partition function of the conditional random field $\Xv_{A(\ell)}$ knowing $\Yv_{A(\ell)}=\y_{A(\ell)}$ and $\Xv_{B(\ell)}=\tilde{\x}_{B(\ell)}$. The latter has shown better performances for estimating the number of latent component $K$ as well as for the selection of the underlying dependency structure $\G$. The criterion nonetheless lack some theoretical support and is limited to regular lattices.

BLIC is to a certain extent related to the RDA approximations of partition functions proposed by \cite{friel2009}. Indeed, let $Z(\btheta,\G)$ and $Z(\bpsi,\G)$ denote the respective normalizing constants of the latent and the conditional fields. Starting from the Bayes formula,
BIC expression turns into
\[
\BIC(m)= -2\frac{\log Z\left(\bpsi,\G\right)}{\log Z\left(\btheta,\G\right)} + d_m\log(n).
\]
Similarly to \cite{friel2009}, \cite{stoehr2016} approximate the intractable normalising constants by a product of tractable normalising constant defined on contiguous sub-lattices. Looking at the issue of estimating the partition function instead of estimating the Gibbs distribution has also been explored by \cite{forbes2003}. They propose to use a first order approximation of the partition function arising from mean field theory. \cite{forbes2003} argue that the latter is more satisfactory than $\BIC^{\mfl}(m)$ in the sense it is based on a optimal lower bound for the normalising constants contrary to the mean field-like approximations. However that does not ensure better results as regards model selection.

Regarding the question of inferring the number of latent states, one might avocate in favor of the Integrated Completed Likelihood \citep[ICL,][]{biernacki2000}. This opportunity has been explored by \cite{cucala2013} but their complex algorithm cannot be extended easily to other model selection problem such as choosing the dependency structure.

\section{Sampling methods for Markov random fields}
\label{sec:sampling-method}

\subsection{Sampling from Gibbs distribution}
\label{sec:simu-hmrf}

Sampling from a Gibbs distribution can be a daunting task due to the correlation structure on a high dimensional space, and standard Monte Carlo methods are impracticable except for very specific cases. In the Bayesian paradigm, Markov chain Monte Carlo (MCMC) methods have played a dominant role in dealing with such problems, the idea being to generate a Markov chain whose stationary distribution is the distribution of interest. The Ising model is one of these special cases where one can be drawn exactly from the model using coupling from the past \citep{propp1996, mira2001}. Nevertheless, such perfect sampling scheme is often prohibitively expensive or impossible to carry out for other models. We describe below two popular solutions even though they introduce a bias.

\subsubsection{Gibbs sampler}

The Gibbs sampler is a highly popular MCMC algorithm in Bayesian analysis starting with the influential development of \cite{geman1984}. It can be seen as a component-wise Metropolis-Hastings algorithm \citep{metropolis1953, hastings1970} where variables are updated one at a time and for which proposal distributions are the full conditionals themselves, see Algorithm \ref{algo:gibbs-sampler}. It is particularly well suited to Markov random field since by nature the intractable joint distribution is fully determined by the easy to compute conditional distributions.

\begin{algorithm2e}[htbp]
  \caption{Gibbs sampler}
  \label{algo:gibbs-sampler}
  \KwIn{a parameter $\btheta$, a number of iterations $T$}
  \KwOut{a sample $\x$ from the joint distribution $f(\cdot\suchthat\btheta,\G)$}
  \medskip
  
  {\bf Initialization:} draw an arbitrary configuration $\x^{(0)}=\left\lbrace x_1^{(0)},\ldots,x_n^{(0)}\right\rbrace$\;
  \For{$t\leftarrow 1$ \KwTo $T$}{
  	\For{$i\leftarrow 1$ \KwTo $n$}{
  		\textbf{draw} $x_i^{(t)}$ from the full conditional $\pr\left(X_i^{(t)}\suchthat\x_{\N(i)}^{(t-1)},\btheta\right)$\;
  	}
  }
  \textbf{return} the configuration $\x^{(T)}$
\end{algorithm2e}
Under the irreducibility assumption, the chain converges to the target distribution $f(\cdot\suchthat\btheta,\G)$, see for example \cite[][\textit{Theorem A}]{geman1984}. Note the order in which the components are updated in Algorithm \ref{algo:gibbs-sampler} does not make much difference as long as every site is visited. Hence it can be deterministically or randomly modified, especially to avoid possible bottlenecks when visiting the configuration space. A synchronous version is nonetheless unavailable since updating the sites merely at the end of cycle $t$ would lead to incorrect limiting distribution.

We should mention here that Gibbs sampler faces some well known difficulties when it is applied to the Ising or Potts model. The Markov chain mixes slowly, namely long range interactions require many iterations to be taken into account, such that switching the color of a large homogeneous area is of low probability even if the distribution of the colors is exchangeable. This peculiarity is even worse when the parameter $\beta$ is above the critical value of the phase transition, the Gibbs distribution being severely multi-modal (each mode corresponding to a single color configuration). \cite{liu1996} proposed a modification of the Gibbs sampler that overcome these drawbacks with a faster rate of convergence. Note also that in the context of Gaussian Markov random field some efficient algorithm have been proposed like the fast sampling procedure of \cite{rue2001}.

\subsubsection{Auxiliary variables and Swendsen-Wang algorithm}

\begin{figure}[t]
\centering
\begin{minipage}[t]{7.cm}
\centering
\hspace*{0.9cm}\includegraphics[width=6.7cm] {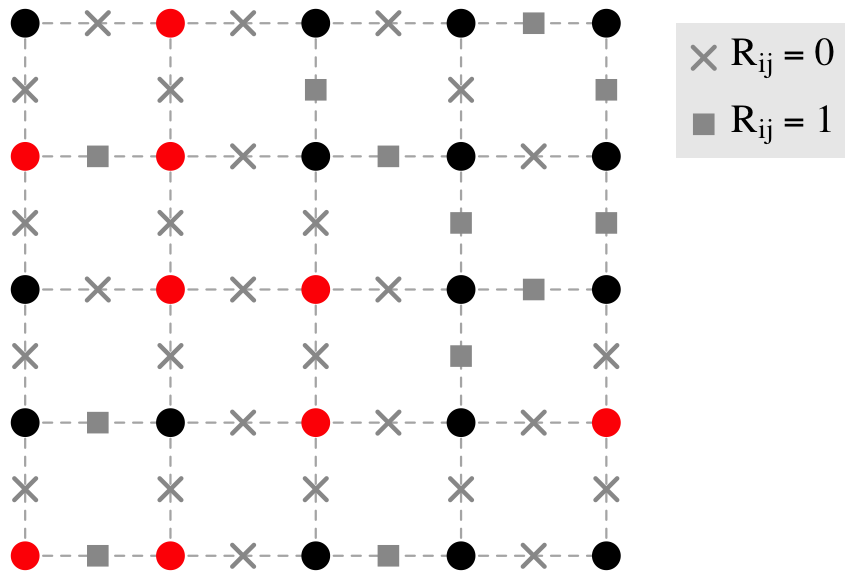}\\
\hspace*{-0.3cm}(a)
\end{minipage}
\begin{minipage}[t]{7.cm}
\centering
\hspace*{0.4cm}\includegraphics[width=6.7cm] {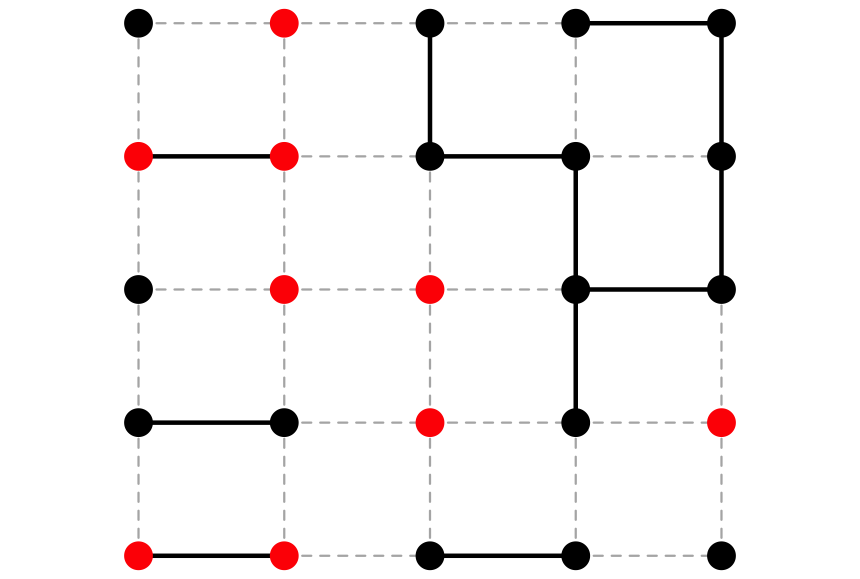}\\
\hspace*{0.4cm}(b)
\end{minipage}
\caption{\small Auxiliary variables and subgraph illustrations for the Swendsen-Wang algorithm. (a) Example of auxiliary variables $U_{ij}$ for a 2-states Potts model configuration on the first order square lattice. (b) Subgraph $\Gamma(\G_4,\x)$ of the first order lattice $\G_4$ induced by the auxiliary variables $U_{ij}$.}
\label{fig:SW}
\end{figure}

An appealing alternative to bypass slow mixing issues of the Gibbs sampler is the Swendsen-Wang algorithm \citep{sw1987} originally designed to speed up simulation of Potts model close to the phase transition. Swendsen-Wang algorithm iterates two steps : a clustering step and a swapping step (see Algorithm \ref{algo:SW}), in order to incorporate simultaneous updates of large homogeneous regions \citep[\textit{e.g., }][]{besag1993}. The clustering step relies on auxiliary random variables which aim at decoupling the complex dependence structure between the component of $\x$ and yield a partition of sites into single-valued clusters or connected components. The method set binary (0-1) conditionally independent auxiliary variables $U_{ij}$ which satisfy
\[
\pr\left(U_{ij}=1\suchthat \x\right) = 
\left\{ 
	\begin{array}{l l}
  	1-\exp\left(\beta_{ij}\ind\{x_i=x_j\}\right)=p_{ij} & \quad \text{if  } \ivj,\\
  	0 & \quad \text{otherwise}\\ 
  	\end{array} 
\right. 
\]
with $\beta_{ij}\geq 0$ so that $p_{ij}$ takes value between 0 and 1. The latter then represents the probability to keep an egde between neighbouring sites in $\G$. Auxiliary variables $U_{ij}$ induce on $\x$ a subgraph $\Gamma(\G,\x)$ of the dependency graph $\G$, namely the undirected graph made of edges of $\G$ for which $U_{ij}=1$, see Figure \ref{fig:SW}. During the swapping step, each cluster $\mathcal{C}$ of $\Gamma(\G,\x)$ is assigned to a new state $k$ with probability
$
\pr\left(\Xv_{\mathcal{C}}=k\right) \propto \exp\left\lbrace\sum_{i\in\mathcal{C}}\alpha_k\right\rbrace.
$
For the special but important case where $\alpha = 0$, new possible states are equally likely. Also for large values of $\beta$, the algorithm manages to switch colors of wide areas, achieving a better cover of the configuration space.

For the original proof of convergence, we refer the reader to \cite{sw1987} and for further discussion see for example \cite{besag1993}. Whilst the ability to change large set of variables in one step seems to be a significant advantage, this can be marred by a slow mixing time, namely exponential in $n$ \citep{gore1999}. The mixing time of the algorithm is polynomial in $n$ for Ising or Potts models with respect to the graphs $\G_4$ and $\G_8$ but only for small enough value of $\beta$ \citep{cooper1999}. This was proved independently by \cite{huber2003} who also derive a diagnostic tool for the convergence of the algorithm to its invariant distribution,
namely using a coupling from the past procedure.

\begin{algorithm2e}[t]
  \caption{Swendsen-Wang algorithm}
  \label{algo:SW}
  \KwIn{a parameter $\btheta$, a number of iterations $T$}
  \KwOut{a sample $\x$ from the joint distribution $f(\cdot\suchthat\btheta,\G)$}
  \medskip
  
  {\bf Initialization:} draw an arbitrary configuration $\x^{(0)}=\left\lbrace x_1^{(0)},\ldots,x_n^{(0)}\right\rbrace$\;
  \For{$t\leftarrow 1$ \KwTo $T$}{
  {\bf Clustering step:}  turn off edges of $\G$ with probability $\exp\left(\beta_{ij}\ind\{x_i^{(t)}=x_j^{(t)}\}\right)$
  {\bf Swapping step:} assign a new state $k$ to each connected component $\mathcal{C}$ of $\Gamma\left(\G,\x^{(t)}\right)$ with probability 
  $\pr\left(\Xv_{\mathcal{C}}^{(t)}=k\right) \propto \exp\left\lbrace\sum_{i\in\mathcal{C}}\alpha_k\right\rbrace$\;
  }
  \textbf{return} the configuration $\x^{(T)}$
\end{algorithm2e}

The algorithm can be extended to other Markov random field or models \citep[\textit{e.g., }][]{edwards1988, wolff1989, higdon1998, barbu2005} but is then not necessarily efficient. In particular, it is not well suited for latent process. The bound to the data corresponds to a non-homogeneous external field that slows down the computation since the clustering step does not make a use of the data. A solution that might be effective is the partial decoupling of \cite{higdon1993, higdon1998}. More recently, \cite{barbu2005} make a move from the data augmentation interpretation to a Metropolis-Hastings perspective in order to generalize the algorithm to arbitrary probabilities on graphs. Nevertheless, it is not straightforward to bound the Markov chain of such modifications and mixing properties are still an open question despite good results in numerical experiments.


\subsection{Monte Carlo maximum likelihood estimator} 

The use of Monte-Carlo techniques in preference to pseudolikelihood to compute maximum likelihood estimates has been especially highlighted by \cite{geyer1992}. Let assume $\pot_{\btheta,\G}$ is continuously differentiable on $\bTheta$. The gradient of the log-density \eqref{eqn:gibbs} can be written as
\[
\grad_{\btheta}\log f(\x\suchthat\btheta,\G) = -\grad_{\btheta}\pot_{\btheta,\G}(\x) - \grad_{\btheta}\log Z(\btheta,\G).
\]
Despite closed form is out of reach, forward-simulations from the likelihood taken at each leapfrog step can be used to provide a Monte Carlo estimate of the gradient, using the following identity,
\begin{align}
\grad_{\btheta}\log Z(\btheta,\G) & = \frac{1}{Z(\btheta,\G)}\grad_{\btheta} Z(\btheta,\G) \nonumber\\
& = \frac{1}{Z(\btheta,\G)}\grad_{\btheta} \int_{\X}\exp\left\lbrace \pot_{\btheta,\G}(\x)\right\rbrace\mu(\dd\x) \nonumber\\
& = \int_{\X} \grad_{\btheta}\pot_{\btheta,\G}(\x)\frac{\exp\left\lbrace \pot_{\btheta,\G}(\x)\right\rbrace}{Z(\btheta,\G)}\mu(\dd\x) \nonumber\\
& = \esp_{\btheta}\left\lbrace \grad_{\btheta}\pot_{\btheta,\G}(\Xv)\right\rbrace.
\label{eqn:z-as-expect}
\end{align}
So far, we have only assumed that $\pot_{\btheta,\G}$ is continuously differentiable on $\bTheta$. However this identity holds under regularity conditions which allow one 
to switch the derivative and integral operators (the domain $\X$ of $\Xv$ is assumed to be independent of $\btheta$) and under the assumption that 
$\grad_{\btheta}\pot_{\btheta,\G}(\Xv)$ is integrable with respect to $f(\x\suchthat\btheta,\G)\mu(\dd\x)$. When the function $\pot_{\btheta}$ linearly depends on the vector of parameters $\btheta$, that is
\[
\pot_{\btheta,\G}(\x) = -\btheta^{T}\Ss(\x),
\]
where $\Ss(\x)=(s_1(\x),\ldots,s_d(\x))$ is a vector of sufficient statistics, expectation \eqref{eqn:z-as-expect} is simply the first moment of the statistics with respect to $f(\x\suchthat\btheta,\G)\mu(\dd\x)$. It is also possible to show that the log-density is concave as its Hessian matrix is the second moment of the statistics with respect to $f(\x\suchthat\btheta,\G)\mu(\dd\x)$:
\[
\grad^{2}_{\btheta} \log f(\x\suchthat\btheta,\G) = -\cov_{\btheta}\left\lbrace \Ss(\Xv)\right\rbrace.
\]
The maximum likelihood estimator $\hat{\btheta}_{\mle}$ is then the unique zero of the score function $\grad_{\btheta} \log f(\x\suchthat\btheta,\G)$ and it satisfies
\begin{equation}
\Ss(\x) - \esp_{\hat{\btheta}_{\text{MLE}}}\left\lbrace \Ss(\Xv)\right\rbrace = 0.
\label{eqn:zero-score}
\end{equation}
Hence a solution to solve problem \eqref{eqn:mle} is to resort to stochastic approximations on the basis of equation \eqref{eqn:zero-score} \citep[\textit{e.g.,}][]{younes1988, descombes1999}. \cite{younes1988} sets a stochastic gradient algorithm converging under mild conditions. At each iteration $t$, the algorithm takes the direction of the gradient estimated by the value of the statistic function for one realisation of the random field $\x^{(t)}$, namely
\[
\btheta^{(t+1)} = \btheta^{(t)} + \frac{\delta}{t+1}\left\lbrace\Ss(\x)-\Ss\left(\x^{(t)}\right)\right\rbrace,
\]
where $\delta$ is a user-defined threshold. \cite{younes1988} provides theoretical conditions on $\delta$ which guarantee the convergence of the algorithm. However such theoretical value for $\delta$ yields in practice a step size too small to ensure the convergence to be achieved in reasonable amount of time. This can be overcome by controlling the probability of non-convergence, see the original paper by \cite{younes1988} for discussion. Another approach to compute the maximum likelihood estimation is to use direct Monte Carlo calculation of the likelihood such as the MCMC algorithm of \cite{geyer1992}. The convergence in probability of the latter toward the maximum likelihood estimator is proven for a wide range of models including Markov random fields. Following that work, \cite{descombes1999} derive also a stochastic algorithm that, as opposed to \cite{younes1988}, takes into account the distance to the maximum likelihood estimator using importance sampling.



\subsection{Computing posterior distributions}
\label{sec:param-posterior}

Markov Chain Monte Carlo allow to asymptotically sample from analytically intractable posterior distribution $\pi$. It provides a very general framework to allow estimation of functionals of the form
\[
\int_{\bTheta} g(\btheta)\pi(\mathrm{d}\btheta),
\]
for some function $g$ by generating a Markov chain $\left(\btheta_n\right)_{n\in\mathbb{N}}$ with transition kernel $P$ which leaves $\pi$ invariant. 
The empirical distribution so obtained leads to the following approximation
\[
\int_{\bTheta} g(\btheta)\pi(\mathrm{d}\btheta) \approx \frac{1}{N}\sum_{n = 1}^N g\left(\btheta_i \right).
\]
While some Bayesian estimators can be efficiently estimated with such methods via the empirical distribution, the intractability of the likelihood model in \eqref{eqn:gibbs} implies, in particular, that the standard MCMC toolbox is infeasible. Indeed, we are concerned with the situation where the un-normalised posterior distribution, the right-hand-side of (\ref{eqn:posterior}) is intractable. This complication results in what is often termed a doubly-intractable
posterior distribution, since the posterior distribution itself is normalised by the evidence (or marginal likelihood) which is typically also intractable. For instance, proposing to move from $(\btheta)$ to $(\btheta')$ in a standard Metropolis-Hastings requires the computation of the unknown normalising constants, $Z(\btheta,\G)$ and $Z(\btheta',\G)$,
\[
\rho(\btheta,\btheta') = 1 \wedge
\frac{
	Z\left(\btheta,\G\right)
}
{
	Z\left(\btheta',\G\right)
} 
\frac{
	q\left(\x\suchthat\btheta',\G\right)\nu(\btheta\mid\btheta')\pi(\btheta')
}
{
	q\left(\x\suchthat\btheta,\G\right)\nu(\btheta'\mid\btheta)\pi(\btheta)
},
\]
where $\nu(\btheta' \mid \btheta)$ denotes the proposal distribution to move from $\btheta$ to $\btheta'$. A solution, while being time consuming, is to estimate the ratio of the partition functions using path sampling \citep{gelman1998}. Starting from equation \eqref{eqn:z-as-expect}, the path sampling identity writes as 
\[
\log\left\lbrace\frac{Z\left(\btheta_0,\G\right)}{Z\left(\btheta_1,\G\right)} \right\rbrace = \int_{\btheta_0}^{\btheta_1} \esp_{\btheta}\lbrace \grad_{\btheta}\pot_{\btheta,\G}(\Xv)\rbrace\mathrm{d}\btheta. 
\]
Hence the ratio of the two normalizing constants can be evaluated with numerical integration. For practical purpose, this approach can barely be recommended within a Metropolis-Hastings scheme since each iteration would require to compute a new ratio.

The exchange algorithm \citep{murray2006} is a popular MCMC methods to allow sampling from doubly-intractable 
distributions. 
The exchange algorithm samples from an augmented distribution
\[
\pi(\btheta', \btheta, \mathbf{u}\knowing \x) \propto f(\x\suchthat\btheta,\G)\pi(\btheta)\nu(\btheta'\knowing\btheta)f(\mathbf{u}\suchthat\btheta',\G).
\] 
whose marginal distribution in $\btheta$ is the posterior distribution of interest. It extends an idea introduced by \cite{moller2006}. The proposal of \cite{moller2006} consists in including an auxiliary variable $\mathbf{U}$ whose density is the intractable likelihood itself. It follows a method based on single point importance sampling approximations of the partition functions $Z(\btheta,\G)$ and $Z(\btheta',\G)$. \cite{murray2006} develop this work further by directly estimating the ratio $Z(\btheta,\G)/Z(\btheta',\G)$ instead of using previous single point estimates.
This leads to a clever algorithm to sample from the above augmented 
distribution,  where it turns out that the ratio of intractable normalising constants drops out of the acceptance probability
\[
\rho\left( \btheta,\btheta',\mathbf{u}\right) = 1\wedge
\frac{\cancel{\textcolor{red}{Z(\btheta,\G)}}}{\cancel{\textcolor{red}{Z(\btheta',\G)}}}
\frac{\cancel{\textcolor{red}{Z(\btheta',\G)}} q(\mathbf{u}\suchthat\btheta,\G)}{q(\mathbf{u}\suchthat\btheta',\G)\cancel{\textcolor{red}{Z(\btheta,\G)}}}
\frac{q(\x\suchthat\btheta',\G)\nu(\btheta\knowing\btheta')\pi(\btheta')}{q(\x\suchthat\btheta,\G)\nu(\btheta'\knowing\btheta)\pi(\btheta)}
\]
\cite{murray2006} point out that the fraction $q(\mathbf{u}\suchthat\btheta,\G)/q(\mathbf{u}\suchthat\btheta',\G)$ which appears above, can be considered as an single sample importance
estimator of $Z(\btheta,\G)/Z(\btheta',\G)$ since it holds that
\begin{equation}
\esp_{\btheta'}\left\lbrace \frac{ q(\mathbf{U}\suchthat\btheta,\G)}{q(\mathbf{U}\suchthat\btheta',\G)}\right\rbrace = \frac{Z(\btheta)}{Z(\btheta')},
\label{eqn:z-ratio-is}
\end{equation}
where $\esp_{\btheta'}$ is the expectation with respect to $\mathbf{U}\sim f(\cdot\suchthat\btheta',\G)$. In fact \cite{alquier2016}, consider a generalised exchange algorithm based, 
at each step of the algorithm, on an improved unbiased estimate of $Z(\btheta)/Z(\btheta')$ including multiple auxiliary draws with respect to the proposed 
parameter, namely,
\begin{equation}
\widehat{\frac{Z\left(\btheta,\G\right)}{Z\left(\btheta',\G\right)}} = \frac{1}{N}\sum_{n=1}^N \frac{ q(\mathbf{u}^{(n)}\suchthat\btheta,\G)}{q(\mathbf{u}^{(n)}\suchthat\btheta',\G)},
\label{eqn:ISE-ratio}
\end{equation}
where the auxiliary variables $\left\lbrace\mathbf{u}^{(1)},\ldots,\mathbf{u}^{(N)}\right\rbrace$ are drawn from $f(\cdot\suchthat\btheta',\G)$. However this so-called noisy exchange 
algorithm no longer leaves the target distribution invariant, nevertheless it is possible to provide convergence guarantees that the resulting Markov chain is close in 
some sense to the target distribution. Following the argument by \cite{everitt2012}, any algorithm producing an unbiased estimate of the normalizing constant can thus be used in place of the importance sampling approximation and will lead to a valid procedure. An alternative to previous methods presented but neglected so far in the literature is Russian Roulette sampling \citep{lyne2015} which can be used to get an unbiased estimate of $1/Z(\btheta)$. 
The idea to apply MCMC methods to situation where the target distribution can be estimated without bias by using an auxiliary variable construction has appeared in the \textit{generalized importance Metropolis-Hasting} of \cite{beaumont2003} and has then been extented by \cite{andrieu2009}. This brings another justification to the aforementioned methods and possible improvement from sequential Monte Carlo literature \citep[\eg][]{andrieu2010}.

\begin{algorithm2e}[!t]
  \caption{Exchange algorithm}
  \label{algo:exchange}
  \KwIn{an initial guess $(\btheta^{(0)}, \btheta'^{(0)}, \mathbf{u}^{(0)})$, a number of iterations $T$}
  \KwOut{a sample 
  drawn from the augmented distribution $\pi\left(\btheta,\btheta',\mathbf{u}\suchthat\x\right)$}
  \medskip
  
  \setstretch{1.35}
  \For{$t\leftarrow 1$ \KwTo $T$}{
  {\bf draw} $\btheta'$ from $\nu(\cdot\suchthat\btheta^{(t-1)})$\;
  {\bf draw} $\mathbf{u}$ from $f\left(\cdot\suchthat\btheta,\G\right)$\;
  {\bf compute} the Metropolis-Hastings acceptance ratio
\[
\rho\left( \btheta^{(t-1)},\btheta',\mathbf{u}\right) = 1\wedge
\frac{q(\mathbf{u}\suchthat\btheta^{(t-1)},\G)}{q(\mathbf{u}\suchthat\btheta',\G)}
\frac{q(\x\suchthat\btheta',\G)\nu(\btheta^{(t-1)}\knowing\btheta')\pi(\btheta')}{q(\x\suchthat\btheta^{(t-1)},\G)\nu(\btheta'\knowing\btheta^{(t-1)})\pi(\btheta)}
\]
  {\bf Exchange move:} set $(\btheta^{(t)}, \btheta'^{(t)}, \mathbf{u}^{(t)}) = (\btheta',\btheta^{(t-1)},\mathbf{u})$ with probability $\rho( \btheta^{(t-1)},\btheta',\mathbf{u})$\;
  {\bf Otherwise} set $(\btheta^{(t)}, \btheta'^{(t)}, \mathbf{u}^{(t)}) = (\btheta^{(t-1)},\btheta'^{(t-1)},\mathbf{u}^{(t-1)})$\;
  }
  \textbf{return} $\left\lbrace \left(\btheta^{(t)}, \btheta'^{(t)}, \mathbf{u}^{(t)}\right) \right\rbrace_{t=1}^T$
\end{algorithm2e}

While the motivation is quite similar, the exchange algorithm is more convenient to implement whilst outperforming the single auxiliary variable method (SAVM) of \cite{moller2006}. Indeed, SAVM requires to choose the conditional distribution $q(\mathbf{u}\mid\x,\btheta)$ which makes it difficult to calibrate. For instance, \cite{cucala2009} stress out that a suitable choice is paramount and may significantly affect the performances of the algorithm. 

A practical difficulty of implementing the exchange algorithm is the requirement to sample $\mathbf{u}$ from Gibbs distribution  $f(\cdot\suchthat\btheta,\G)$ to guarantee a valid MCMC scheme. In all generality, it is impossible or prohibitively expensive to carry out such perfect sampling. \cite{everitt2012} has provided convergence results when one uses instead the final draw from a Gibbs sampler with stationary distribution $f(\cdot\suchthat\btheta,\G)$ as an approximate realisation. \cite{everitt2012} has notably pointed out that solely few iterations of the sampler are necessary. This approach has shown good performances in practice \citep[\eg,][]{cucala2009, caimo2011}.

While the application of the exchange algorithm is straightforward for Bayesian parameter inference of a fully observed Markov random field, some work have been devoted to the use of the exchange algorithm for hidden Markov random fields such as the exchange marginal particle MCMC of \cite{everitt2012} or the estimation procedure in \cite{cucala2013}. Though these methods produce accurate results they inherit the drawback of the exchange algorithm.

\subsection{ABC model selection}
\label{subsec:abc-model}

Approximate Bayesian computation (ABC) has generated much activity in the literature recently as it offers a way to circumvent the difficulties of models which are intractable but can be simulated from. We refer the reader to \cite{marin2012} and the references therein for a comprehensive overview on the method. When performing parameter estimation, the method is particularly well suited for problems where the likelihood function does not admit an algebraic form, a situation where MCMC methods are at a loss. We believe that the benefit of ABC for parameter estimation of a Gibbs random field is questionable. Hence we solely focus on model selection in this part.

To approximate $\widehat{m}_\text{MAP}$, ABC starts by simulating numerous triplets $(m,\btheta_m,\y)$ from the joint Bayesian model, see Algorithm \ref{algo:abc2}. Afterwards, the algorithm mimics the Bayes classifier \eqref{eq:bayes-classifier}: it approximates the posterior probabilities by the frequency of each model number associated with simulated $\y$'s in a neighbourhood of $\yobs$. In a metric space $(\Y,\rho)$, this neighbourhood is define by the ball $\mathcal{B}(\epsilon,\yobs)$ of radius $\epsilon$ centered at $\yobs$. If required, we can eventually predict the best model with the most frequent model in the neighbourhood, or, in other words, take the final decision by plugging in \eqref{eq:bayes-classifier} the approximations of the posterior probabilities. However such a naive implementation is typically infeasible as the data usually lies in a space of high dimension and the algorithm faces the curse of dimensionality. Put in other words, sample dataset in the neighbourhood of $\y$ occurs with an prohibitively low probability. The ABC algorithm performs therefore a (non linear) projection of the observed and simulated datasets onto some Euclidean space of reasonable dimension $d$ via a function $s$, composed of summary statistics $\Ss(\cdot) =\big\lbrace s_1(\cdot),\ldots,s_M(\cdot)\big\rbrace$ that are the concatenation of the summary statistics of each models with cancellation of possible replicates.. The neighbourhood of $\yobs$ is thus defined as simulations whose distances to the observation measured in terms of summary statistics, \ie $\rho\left\lbrace\Ss(\y),\Ss(\yobs)\right\rbrace$, fall below a threshold $\step$. The accepted particles $(m^{(t)}, \y^{(t)})$ at the end of Algorithm \ref{algo:abc2} are distributed according to the pseudo-target $\pi(m\mid \rho\left(\Ss(\y),\Ss(\yobs)\right)\leq\epsilon)$ and the estimate of the posterior model distribution is given by
\[
\hat{\pi}_\epsilon\left(m\suchthat\yobs\right) = \frac{\sum \ind\left\lbrace m^{(t)} = m,  \rho\left(\Ss\left(\y^{(t)}\right),\Ss\left(\yobs\right)\right)\leq\epsilon\right\rbrace}{\sum \ind\left\lbrace \rho\left(\Ss\left(\y^{(t)}\right),\Ss\left(\yobs\right)\right)\leq\epsilon\right\rbrace}.
\] 
ABC hence presents two level of approximations arising from the size of the neighbourhood $\epsilon$ and the introduction of summary statistics.

\begin{algorithm2e}
  \caption{ABC model choice algorithm}
  \label{algo:abc2}
  \KwIn{an observation $\yobs$, summary statistics $\Ss$, a number of iterations $T$, an empirical quantile of the distance $T_\epsilon$}
  \KwOut{a sample from the approximated target of $\pi_\epsilon\left(\cdot\mid\Ss(\yobs)\right)$}
  \medskip
  
  \For{$t\leftarrow 1$ \KwTo $T$}{
  {\bf draw} $m$ from $\pi$\;
  {\bf draw} $\bpsi$ from $\pi_m$\;
  {\bf draw} $\y$ from $f_m(\cdot\mid\bpsi)$\;
   {\bf compute} $\Ss(\y)$\;
  {\bf save} $\left\lbrace m^{(t)},\btheta^{(t)},\Ss\left(\y^{(t)}\right)\right\rbrace \leftarrow \left\lbrace m, \bpsi,\Ss\left(\y\right)\right\rbrace$\; 
}
 {\bf sort} the replicates according to the distance $\rho\left(\Ss\left(\y^{(t)}\right),\Ss\left(\yobs\right)\right)$\;
   {\bf keep} the $T_\epsilon$ first replicates\; 
  
  \textbf{return} the sample of accepted particles
\end{algorithm2e}

The choice of such summary statistics presents major difficulties that have been especially highlighted for model choice \citep{robert2011, didelot2011}. When the summary statistics are not sufficient for the model choice problem, \cite{didelot2011} and \cite{robert2011} found that the above probability can greatly differ from the genuine $\pi(m\mid\yobs)$. Model selection between fully observed Markov random fields whose energy function is of the form $\pot_{\btheta,\G}(\y) = \btheta^{T}s(\y)$ is a surprising example for which ABC is consistent. Indeed \cite{grelaud2009} have pointed out that vector of summary statistics $\Ss(\cdot) =\big\lbrace s_1(\cdot),\ldots,s_M(\cdot)\big\rbrace$ is sufficient for each model but also for the joint parameter across models $(\M,\theta_1,\ldots,\theta_M)$. This allows to sample exactly from the posterior model distribution when $\epsilon=0$. However the fact that the concatenated statistic inherits the sufficiency property from the sufficient statistics of each model is specific to exponential families \citep{didelot2011}. When dealing with model choice between hidden Markov random fields, we fall outside of the exponential families due to the bound to the data. Thus we face the major difficulty outlined by \cite{robert2011}: it is almost impossible to build a sufficient statistic of reasonable dimension, \ie, of dimension much lower than the dimension of $\X$.

Beyond the seldom situations where sufficient statistics exist and are explicitly known, \citet{marin2014} provide conditions which ensure the consistency of ABC model choice but the latter are difficult, if not impossible, to check in practice. To answer the absence of sufficient statistics as well as aforementioned theoretical conditions, very few has been accomplished in the context of ABC model choice. One solution would be to rely on the approach of \citet{prangle2014}. The statistics $\Ss(y)$ reconstructed by \citet{prangle2014} have good theoretical properties (those are the posterior probabilities of the models in competition). Nevertheless the methods requires a pilot ABC run which is time consuming and can lead to poor approximations. Alternatively, \cite{stoehr2015} have proposed to use summary statistics based on connected components of induced graphs in order to select between dependency structures of hidden Markov random fields. Beside, the specific result on Markov random fields, they derive an adaptive scheme to select the most informative set of statistics based on a local error rate. The main drawback of the method is that its use is limited to low dimensional vector of statistics. To overcome this issue, the summary statistics proposed by \cite{stoehr2015} could be used among others within the ABC random forest procedure of \cite{pudlo2015}.



\end{document}